\documentclass{aa}

\usepackage{graphicx}
\usepackage{amsmath} 
\usepackage{amssymb}
\usepackage{lscape}
\usepackage{longtable,lscape}
\usepackage{aalongtable,lscape}
\usepackage[sort&compress]{natbib}
\usepackage{natbib}
\usepackage{pifont}
\usepackage{dingbat}

  \bibpunct{(}{)}{;}{a}{}{,}

\usepackage{txfonts}

\begin{document}

   \title{A WFI survey in the Chamaeleon~II dark cloud\thanks{Based on observations carried out at the European Southern Observatory, La Silla, Chile under proposals 
           numbers 67.C-0225 and 68.C-0311.}}

  \author{L. Spezzi\inst{1} 
   \and J.M. Alcal\'{a}\inst{2} 
   \and A. Frasca\inst{1} 
   \and E. Covino\inst{2} 
   \and D. Gandolfi\inst{3}}

   \offprints{L. Spezzi, \email{lspezzi@oact.inaf.it}}

\institute{INAF - Osservatorio Astrofisico di Catania, via S. Sofia, 78, 95123 Catania, Italy 
\and INAF - Osservatorio Astronomico di Capodimonte, Salita Moiariello, 16, 80131 Napoli, Italy 
\and Dipartimento di Fisica e Astronomia, Universit\`a di Catania, via S. Sofia 78, 95123 Catania, Italy}

   \date{Received ; accepted }

  \abstract
{}
{We present the results of an optical multi-band survey for low-mass 
Pre-Main Sequence (PMS) stars and young Brown Dwarfs (BDs) in the 
Chamaeleon~II (Cha~II) dark cloud. This survey constitutes the complementary 
optical data to the c2d Spitzer Legacy survey in Cha~II.}
{Using the Wide-Field Imager (WFI) at the ESO 2.2\,m telescope, we surveyed 
a sky area of about 1.75 square degrees in Cha~II. The region was observed 
in the $R_{\rm c}$, $I_{\rm c}$ and $z$ broad-bands, in H$\alpha$ and in two 
medium-band filters centered at 856 and 914~nm.  
We select PMS star and young BD candidates using colour-magnitude diagrams 
(CMDs) and theoretical isochrones reproduced \emph{ad-hoc} for the WFI 
at the ESO2.2m telescope system. The selection criteria are also 
reinforced by using the previously known PMS stars in Cha~II to define the 
PMS locus on the CMDs and by investigating the infrared (IR) colours of the 
candidates. By exploiting the WFI intermediate-band photometry we also 
estimate the effective temperature and the level of H$\alpha$ emission of 
the candidates.}
{Our survey, which is one of the largest and deepest optical surveys 
conducted so far in Cha~II, recovered the majority of the PMS 
stars and 10  member candidates of the cloud from previous 
IR surveys. In addition, the survey revealed 10 new potential 
members. 
From our photometric characterisation, we estimate 
that some 50\% of the 20 candidates will result in true Cha~II members. 
Based on our temperature estimates, we conclude that several of these 
objects are expected to be sub-stellar and give a first estimate 
of the fraction of sub-stellar objects.}
{}


   \maketitle

\section{Introduction}
\label{sec:Intro}

Recent investigations in star-forming regions (SFRs) have pointed 
out that the fraction of young Brown Dwarfs (BDs) relative to 
low-mass and more massive Pre-Main Sequence (PMS) stars may vary 
significantly among different SFRs \citep{Kro02}. In particular, the 
fraction is different in T and in OB associations \citep{Hil00, Luh00,
Bri02, Pre03, Mue03}.  Different environments may give rise to 
different initial conditions for star and planet formation and,  
hence, to differences in the observed spectrum of masses \citep{Kro01,Kro02}. 
This may have strong consequences on the Initial Mass Function (IMF), 
in particular in the sub-stellar domain.
It has also been proposed that BDs may form as members of small 
groups of objects, that may be ejected by dynamical interactions 
before they can grow to stellar masses \citep{Rei01}; 
hence, many low-mass and sub-stellar objects might have escaped 
detection in surveys that concentrate sharply in the cores of SFRs. 
Thus, one possible reason for the 
differences in the fraction of young sub-stellar objects in 
T and OB associations can in principle be ascribed to the photometric 
and spatial incompleteness of the imaging surveys.
Spatially complete deep imaging surveys are thus crucial in order 
to single out low-mass star and BD candidates to be investigated by 
follow-up spectroscopy. Only then, problems like mass segregation 
in SFRs and the low-mass end of the IMF can be addressed.

In this paper, optical wide-field imaging observations 
in the Chamaeleon~II dark cloud (hereafter Cha~II), complemented 
with $JHK$ photometry from 2MASS \citep{Cut03}, are used to search 
for low-mass PMS stars and young BDs. Our survey covers almost 
2 square degrees in the Cha~II cloud.

Because of its proximity to the Sun \citep[d$\approx$178~pc,][]{Whi97}, 
young age \citep[0.1-10 Myr,][]{Hug92} and relatively high galactic 
latitude \citep[b$\approx$$-$15~deg,][]{Sch91}, which decreases the 
effects of contamination by background stars, the Cha~II dark cloud 
is particularly well-suited for studies of low-mass PMS stars and 
young BDs. It is indeed characterised by the presence of objects 
with H$\alpha$ emission \citep{Hug92, Har93}, as well as of 
embedded Class-I and Class-II IR sources \citep{Whi91, Pru92, Lar98} 
and X-rays sources \citep{Alc00}.  

Investigations in the near-IR by the DENIS survey revealed 
several candidate young BDs in Cha\,II \citep{Vuo01}; 
however, a spectroscopic follow-up by \citet{Bar04}, though 
revealing the least massive classical T\,Tauri star in the 
cloud, failed in confirming the young BD candidates. \citet{Per03} 
performed ISOCAM observations and IR spectroscopy of several 
objects in the core of Cha~II and found a number of sources 
with IR excess. Their most promising candidate, ISO-CHA\,II\,13, 
was confirmed recently as the first BD in the region known to be
surrounded by a disk \citep{Alc06}. 
An optical wide-field imaging survey by \citet{Lop05} 
proposed two young BD candidates. However, that survey 
covered only about 10\% of the cloud area. 

The Cha\,II cloud has also been included in the Spitzer 
Legacy survey "From Molecular Cores to Planet Forming Disks" 
or c2d \citep{Eva03} as a test case of a cloud with moderate 
star formation activity. Results of the c2d survey in Cha~II 
have been published recently by \citet{You05}, \citet{All06} and 
\citet{Por06} .
The data presented in this paper constitute the optical ancillary 
data for the c2d survey in Cha~II. As such, they are part 
of a multi-wavelength study of the Cha~II cloud which will be 
presented in a forthcoming paper (Alcal\'a et al., in preparation).

The outline of the paper is as follows. In Sec.~\ref{obsred} the 
observations, data reduction and calibration procedures are described. 
In Sec.~\ref{analysis} we present specific tools for the analysis of the 
photometric data. Sec.~\ref{sel_cand} describes the criteria for the 
selection of PMS star and BD candidates in Cha~II, using the tools 
developed in Sec.~\ref{analysis}. 
The results of the survey and, in particular, the fraction of 
sub-stellar objects estimated in Cha~II are discussed in 
Sec.~\ref{sec:par}. 
Our conclusions are presented in Sec.~\ref{sec:concl} and, 
finally, a few notes on some individual objects are presented 
in Appendix~\ref{comm}.

\section{Observations and data reduction}
\label{obsred}

\subsection{Observations}
\label{obs}

\begin{figure*} 
\centering
\includegraphics[width=18cm,height=17cm]{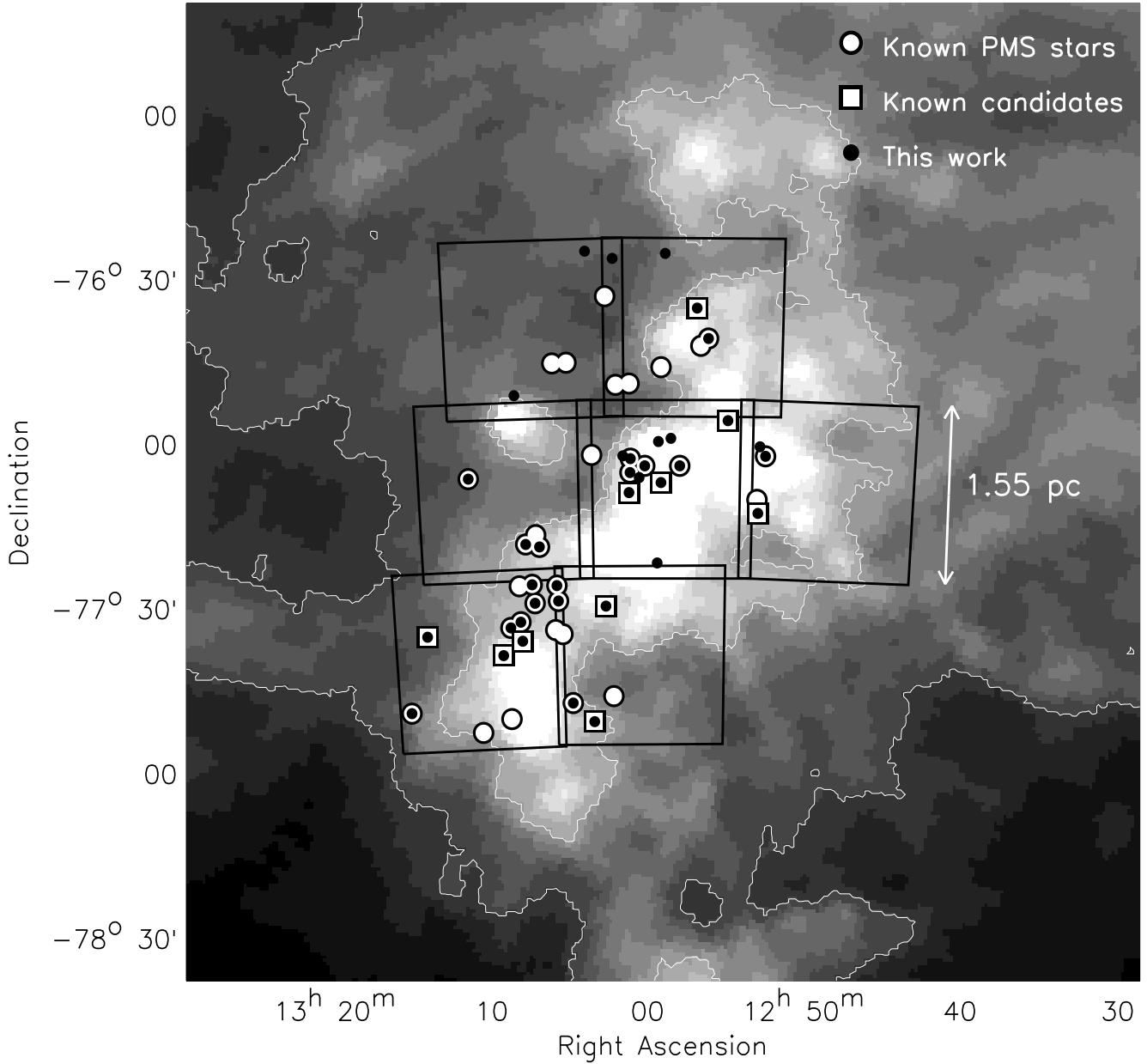}
\caption{IRAS 100$\mu$m dust emission map of the Cha~II dark cloud. 
The contours, from 10 to 40 MJy $\times$ sr$^{-1}$ in steps of 
2.5 MJy $\times$ sr$^{-1}$ are also drawn (white lines). The big 
polygons define the area covered by the seven ESO2.2m+WFI 
pointings. The confirmed members of the cloud and the candidates 
selected in previous surveys are represented with open circles
and squares respectively; the black dots mark the objects 
recovered by our selection criteria (see Sec.~\ref{sel_cand}).
Some 13 previously known PMS stars appear saturated in at least
one of the bands of our survey and this is why they appear as 
open circles only. However, when using photometry from the 
literature, these objects are also recovered from our selection 
criteria (see Sec.~\ref{sel_cand}).}
\label{RaDec}
\end{figure*}

The observations were carried out in two observing runs (27-30 April 2001 
and 20-23 March 2002) using the Wide Field Imaging (WFI) mosaic camera 
attached to the ESO 2.2m telescope at La Silla (Chile). The mosaic 
consists of eight 2k$\times$4k CCDs forming a 8k$\times$8k array with 
a pixel scale of 0.238\arcsec/pix; hence, a single WFI pointing covers a 
sky area of about 30\arcmin$\times$30\arcmin. 
The Cha~II dark cloud has an extention of about 2 square degrees \citep{Hug92}. 
Thus, seven adjacent WFI pointings allowed us to cover about 70\% of the 
cloud area. The distribution on the sky of the WFI pointings is shown in 
Fig.~\ref{RaDec}. The overlap of about 2\arcmin~ between adjacent pointings 
allowed us to check the consistency in the photometry as well as in the 
astrometry.

The observations were performed in the $R_C$, $I_C$ and $z$ broad-bands, 
in two H$\alpha$ filters, narrow ($H\alpha_7$, ${\rm \lambda_c}$=658~nm and 
FWHM=7.4~nm) and wide ($H\alpha_{12}$, ${\rm \lambda_c}$=665~nm and FWHM=12.1~nm), 
and in two intermediate-band filters centered at 865 and 914~nm.  
In order to cover the gaps between the WFI CCDs and to correct for 
moving objects and cosmic ray hits, for every pointing a sequence of 
typically five ditherings was performed in each filter. 
The journal of the observations is presented in Tab.~\ref{tab:obs}. \\

%---------------------------------- RA.vs.DEC-------------------
%-------------------------------------------------------------- 

\begin{table}
\caption[ ]{\label{tab:obs} Journal of the observations. The R.A. (in hh:mm:ss) 
and Dec. (in dd:mm:ss)  of each pointing are indicated between parenthesis. 
Total exposure time (T$_{\rm exp}$), seeing and air mass (X) values  
correspond to the final staked images (see Sec.~\ref{pre_astro}).}
\begin{center}
\scriptsize
\begin{tabular}{cccccc}  
\hline
Field          & Date       &  Filter  & T$_{\rm exp}$ & Seeing & X \\
(RA,DEC)       & (d/m/y)    &          & (min)     & (")    &   \\
\noalign{\medskip}
\hline
\noalign{\medskip}
ChaII\_1$^\dagger$   & 22/03/02 &  Rc/162         & 2$\times$5 & 1.2 & 1.48 \\  
(13:06:18,-76:44:21) & 22/03/02 &  Ic/Iwp         & 2$\times$5 & 1.2 & 1.48 \\  
                     & 22/03/02 &  z+/61          & 2$\times$5 & 1.2 & 1.49 \\  
                     & 21/03/02 &  H$\alpha $/7   & 5$\times$4 & 1.0 & 1.54 \\
                     & 21/03/02 &  H$\alpha $/12  & 3$\times$5 & 1.0 & 1.51 \\ 
\hline
ChaII\_2             & 22/03/02 &  Rc/162	  & 2$\times$5 & 1.2 & 1.55 \\
(12:57:31,-76:44:20) & 22/03/02 &  Ic/Iwp	  & 2$\times$5 & 1.2 & 1.53 \\
   		     & 22/03/02 &  z+/61	  & 2$\times$5 & 1.2 & 1.51 \\  
    		     & 21/03/02 &  H$\alpha$/7    & 5$\times$5 & 1.0 & 1.53 \\
    		     & 21/03/02 &  H$\alpha$/12   & 3$\times$5 & 1.0 & 1.56 \\ 
     		     & 23/03/02 &  856/14	  & 4$\times$5 & 1.0 & 1.50 \\ 
      		     & 23/03/02 &  914/27	  & 5$\times$5 & 1.0 & 1.48 \\ 
\hline
ChaII\_3             & 29/04/01 & Rc/162	  & 10$\times$5& 2.7 & 1.84 \\
(13:08:13,-77:14:27) & 29/04/01 & Ic/Iwp	  & 10$\times$5& 2.7 & 1.69 \\
		     & 29/04/01 & z+/61	          & 10$\times$5& 2.7 & 1.60 \\  
 		     & 21/03/02 & H$\alpha$/7     & 5$\times$5 & 1.2 & 1.64 \\
 		     & 21/03/02 & H$\alpha$/12    & 3$\times$5 & 1.2 & 1.59 \\ 
 		     & 23/03/02 & 856/14	  & 4$\times$5 & 1.2 & 1.51 \\ 
 		     & 23/03/02 & 914/27	  & 5$\times$5 & 1.2 & 1.53 \\ 
\hline 
ChaII\_4             & 22/03/02 & Rc/162          & 2$\times$5 & 1.4 & 1.49 \\ 
(12:59:03,-77:13:58) & 22/03/02 & Ic/Iwp          & 2$\times$5 & 1.4 & 1.50 \\
		     & 22/03/02 & z+/61	          & 2$\times$5 & 1.4 & 1.50 \\  
		     & 21/03/02 & H$\alpha$/7     & 5$\times$5 & 1.2 & 1.50 \\
		     & 21/03/02 & H$\alpha$/12    & 3$\times$4 & 1.2 & 1.51 \\ 
		     & 23/03/02 & 856/14	  & 4$\times$5 & 1.2 & 1.57 \\
		     & 24/03/02 & 914/27	  & 5$\times$5 & 1.2 & 1.54 \\ 
\hline
ChaII\_5             & 22/03/02 & Rc/162	  & 2$\times$5 & 1.2 & 1.60 \\
(12:49:53,-77:13:59) & 22/03/02 & Ic/Iwp	  & 2$\times$5 & 1.2 & 1.62 \\
 		     & 22/03/02 & z+/61	          & 2$\times$5 & 1.2 & 1.66 \\  
 		     & 21/03/02 & H$\alpha$/7     & 5$\times$5 & 1.2 & 1.75 \\
 		     & 21/03/02 & H$\alpha$/12    & 3$\times$5 & 1.2 & 1.66 \\ 
 		     & 23/03/02 & 856/14	  & 4$\times$5 & 1.0 & 1.49 \\
		     & 23/03/02 & 914/27	  & 5$\times$5 & 1.0 & 1.50 \\ 
\hline
ChaII\_6             & 30/04/01 & Rc/162          & 10$\times$6& 2.0 & 1.67 \\ 
(13:10:07,-77:45:08) & 30/04/01 & Ic/Iwp          & 10$\times$6& 2.0 & 1.57 \\ 
		     & 30/04/01 & z+/61	          & 10$\times$6& 2.0 & 1.52 \\  
		     & 21/03/02 & H$\alpha$/7     & 5$\times$5 & 1.4 & 1.61 \\
		     & 21/03/02 & H$\alpha$/12    & 3$\times$5 & 1.4 & 1.66 \\ 
		     & 23/03/02 & 856/14	  & 4$\times$5 & 1.2 & 1.67 \\
 		     & 23/03/02 & 914/27          & 5$\times$5 & 1.2 & 1.74 \\ 
\hline
ChaII\_7             & 22/03/02 &  Rc/162         & 2$\times$4 & 1.2 & 1.78 \\
(13:00:28,-77:44:39) & 22/03/02 &  Ic/Iwp         & 2$\times$5 & 1.2 & 1.74 \\
                     & 22/03/02 &  z+/61          & 2$\times$5 & 1.2 & 1.69 \\  
		     & 21/03/02 &  H$\alpha$/7    & 5$\times$5 & 1.0 & 1.52 \\
		     & 21/03/02 &  H$\alpha$/12   & 3$\times$5 & 1.0 & 1.51 \\ 
		     & 23/03/02 &  856/14	  & 4$\times$5 & 1.0 & 1.58 \\
 		     & 23/03/02 &  914/27	  & 5$\times$5 & 1.0 & 1.62 \\ 
\noalign{\medskip}
\hline
\end{tabular}
\end{center}
$^\dagger$ \footnotesize{Not observed in the 856-nm and 914-nm bands.}
\end{table}

\subsection{Pre-reduction and Astrometry}

\label{pre_astro}

The pre-reduction of the raw images, passing through bias subtraction, 
flat-fielding and correction for fringing, was performed on a 
nightly basis, using the {\em mscred} package under 
IRAF\footnote{IRAF is distributed by the National Optical Astronomy 
Observatories (NOAO).}, following the guidelines 
described in \citet{Alc02}. The resulting pre-reduced images are 
uniform to about 1\%. In order to correct for large-scale 
illumination variations a super-flat was used.

The astrometric calibration and relative flux scaling between ditherings 
were done using the ASTROMETRIX tool\footnote{Available at: http://www.na.astro.it/~radovich.}. 
ASTROMETRIX performs a global astrometric solution which takes into account overlapping
sources falling on adjacent CCDs in different ditherings. For each pointing, the astrometric 
solution was first computed for the $R$-band dithering-set using the USNO-A2.0 
catalogue \citep{Mon98} as a reference. A catalogue of sources was then extracted 
from the re-sampled  $R$-band image and used as reference catalogue for all 
the other bands. Within the global astrometry process, the astrometric 
solution was constrained for each CCD by both the positions from the 
$R$-band catalogue and those from overlapping sources in all the other CCDs. 
The co-addition of the different dithered images for a given filter and 
pointing was performed using the SWARP tool\footnote{Available at:  http://terapix.iap.fr/cplt/oldSite/soft/swarp.}. 
The resulting 8k$\times$8k stacked images were normalised by the exposure 
time. The absolute astrometric precision of our images is about 0.35 arcsec, 
corresponding to the RMS accuracy of the USNO-A2.0 catalogue, while the 
internal RMS, computed from overlapping sources in different exposures, 
is within 0.05 arcsec, indicating the good performance of ASTROMETRIX.

\subsection{The photometric calibration}
\label{photcal}

In order to transform the $R$ and $I$ instrumental magnitudes to the standard
Cousins system, the Landolt standard fields SA~98, SA~101 and SA~107 \citep{Lan92} were 
observed nightly. The standard $R_C$ and $I_C$ magnitudes were determined 
using the transformation equations:

\begin{equation}
R_C=r_0 + c_R \cdot (r_0-i_0)+ZP_R 
\end{equation}

\begin{equation}
I_C=i_0 +c_I \cdot (r_0-i_0)+ZP_I 
\end{equation}

where  $r_0$ and $i_0$ are the instrumental magnitudes corrected for 
atmospheric extinction and $c_R$ and $ZP_R$ and $c_I$ and $ZP_I$ the colour 
terms and zero points for the $R$ and $I$ bands respectively. The standard 
magnitudes from \citet{Ste00} were used in order to determine the zero points 
and colour terms.

The intermediate-band instrumental photometry was transformed to the 
standard AB photometric system following the prescriptions by \citet{Jac87} 
and \citet{Alc02} by using the equation:

\begin{equation}
m_{AB}(\lambda)=m_{0}(\lambda) + ZP_{\lambda}
\end{equation}

where $m_{0}$ is the instrumental magnitude in a given intermediate-band 
filter corrected for atmospheric extinction and $ZP_{\lambda}$ the zero 
point derived by using the spectrophotometric standard stars Hiltner~600, 
LTT~4364 and Eg~274 \citep{Ham92}, observed by us in the same run. 
The resulting magnitudes in the intermediate band filters are in
the AB system.

The nightly calibration coefficients are reported in Tab.~\ref{tab:coeff}.

% --------------------------------------------- Table ----------------------------
\begin{table}
\caption[ ]{\label{tab:coeff} Photometric calibration coefficients 
(atmospheric extinction coefficient, {\it K}, zero point, {\it ZP}, and colour 
terms, {\it c}) for the WFI filters used in this work.}
\begin{center}
\small
\begin{tabular}{ccccc}  
\hline
Filter & Date & {\it K}$^\dagger$ & {\it ZP} & {\it c} \\
\noalign{\medskip}
\hline
\noalign{\medskip}
$R$		& 04/27/2001 & 0.096 & 24.418$\pm$0.004 & -0.038$\pm$0.010  \\
"		& 04/28/2001 & 0.096 & 24.388$\pm$0.001 & -0.056$\pm$0.008  \\
"		& 04/29/2001 & 0.096 & 24.445$\pm$0.004 & -0.023$\pm$0.003  \\
"		& 03/21/2002 & 0.096 & 24.397$\pm$0.003 & -0.030$\pm$0.003  \\
$I$		& 04/27/2001 & 0.082 & 23.413$\pm$0.012 & 0.222$\pm$0.002   \\
"		& 04/28/2001 & 0.082 & 23.399$\pm$0.001 & 0.216$\pm$0.003   \\
"		& 04/29/2001 & 0.082 & 23.431$\pm$0.001 & 0.221$\pm$0.007   \\
"		& 03/21/2002 & 0.082 & 23.335$\pm$0.001 & 0.222$\pm$0.002   \\
$z$		& 04/27/2001 & 0.080 & 21.546$\pm$0.804 & 0.336$\pm$0.311   \\
"		& 04/28/2001 & 0.080 & 21.246$\pm$0.303 & 0.230$\pm$0.143   \\
"		& 04/29/2001 & 0.080 & 21.248$\pm$0.353 & 0.214$\pm$0.140   \\
"		& 03/21/2002 & 0.080 & 21.460$\pm$0.014 & 0.373$\pm$0.001   \\	  		 
$H\alpha_7$	& 03/20/2002 & 0.096 & 21.249$\pm$0.095 &  	/	    \\
$H\alpha_{12}$	& 03/20/2002 & 0.096 & 21.733$\pm$0.133 &  	/  	    \\
$856nm$	        & 03/22/2002 & 0.080 & 20.894$\pm$0.089 &  	/ 	    \\
$914nm$	        & 03/22/2002 & 0.080 & 21.018$\pm$0.082 &  	/ 	    \\
\noalign{\medskip}
\hline
\end{tabular}
\end{center}
$^\dagger$ \footnotesize{The mean atmospheric extinction coefficients for La Silla 
have been adopted.}
\end{table}
%-------------------------------------------------------------------------------------

\subsubsection{The \emph{z}-band photometric calibration}
\label{calZ}

The calibration of the $z$ magnitudes required more effort as
there are no data for standard stars available in the literature. Since we
aimed at determining the colour-magnitude diagrams (CMDs) of the sources in Cha~II (Sec.~\ref{sel_cand}) 
and their spectral energy distributions (SEDs, Sec.~\ref{sec:par}), we needed both 
to tie the $z$ magnitudes with some ``reference'' photometric system and to 
determine the flux at Earth of a star with magnitude $z=0$.
To this aim, we observed the same Landolt standard fields in the $z$ filter. 
In the Landolt SA~98 field we selected the A0-type star SA~98~653, whose 
visual magnitude and colours are reported in  Tab.~\ref{SA98}. This star is 
very well characterised and its absolute spectrophotometry is well determined 
\citep{Gut88}. For this A0-type star we can assume $(I_C-z)=0$, thus defining 
the ``standard'' $z$-magnitude system as that for which the colour ($I_C-z$) 
is zero for A0 type stars. The zero point of this calibration is then 
determined as follows:

\begin{equation}
Zp=z-z_0=I_C-z_0
\end{equation}

where $I_C=9.522$ is the standard magnitude of SA~98~653 in the Cousins 
system and $z_0$ its $z$-band instrumental magnitude corrected for 
atmospheric extinction.
We then applied this zero point correction to the $z_0$ instrumental magnitude 
of all the stars in the Landolt fields. 
In this way we obtained a catalogue of Landolt stars with $z$ magnitudes 
and $(I-z)$ colours in our internal WFI-Cousins system. 
We then used this catalogue to find the least squares solution to the 
equation:

\begin{equation}
z=z_0 + c_z \cdot (i_0-z_0) + ZP_z
\end{equation}

where $z_0$ and $i_0$ are the standard stars instrumental magnitudes 
corrected for atmospheric extinction and $ZP_z$ and $c_z$ the $z$ 
calibration coefficients for the newly defined WFI-Cousins photometric 
system. 
The nightly mean values obtained for $ZP_z$ and $c_z$ are reported 
in Tab.~\ref{tab:coeff}.

%----------------------------------  Table ---------------------------
\begin{table}
\caption[ ]{\label{SA98} Standard photometry of SA~98~653 
(R.A. = 06:52:05, Dec. = $-$00:18:18) from \citet{Ste00}.}
\begin{center}
\begin{tabular}{ccccc}  
\hline
$V$    & $(B-V)$ & $(U-B)$ & $(V-R_C)$ & ($R-I$)$_C$ \\ 
\noalign{\medskip}
\hline
\noalign{\medskip}
9.539  & $-$0.004  & $-$0.097   &  0.007  &  0.008  \\	    
\noalign{\medskip}
\hline
\end{tabular}
\end{center}
\end{table}
%-------------------------------------------------------------------------

Since the absolute spectrophotometry of SA~98~653 is available from 
\citet{Gut88}, we can use this star to obtain the absolute flux calibration 
in the $z$ filter, i.e. to obtain the flux at Earth of a star with $z$=0. 
Following the procedure outlined in Appendix~\ref{fluxZ} we 
derive:

$$
F(z=0) = (8.4 \pm 0.1) \cdot 10^{-10}\ \ {\rm erg\, cm}^{-2}\ {\rm s}^{-1}\ \AA^{-1}
$$ 
$$ 
\hspace{-1.9cm}       = (2608 \pm 31)\ {\rm Jy}
$$ 
 
This value can be used to derive the flux in the $z$-band of the sources 
in our catalogue using their corresponding $z$-band magnitudes.

\subsection{The catalogue extraction}
\label{PSF}

%----------------------------------errors PSF -------------------
\begin{figure} 
\resizebox{\hsize}{!}{\includegraphics[width=15cm,height=25cm]{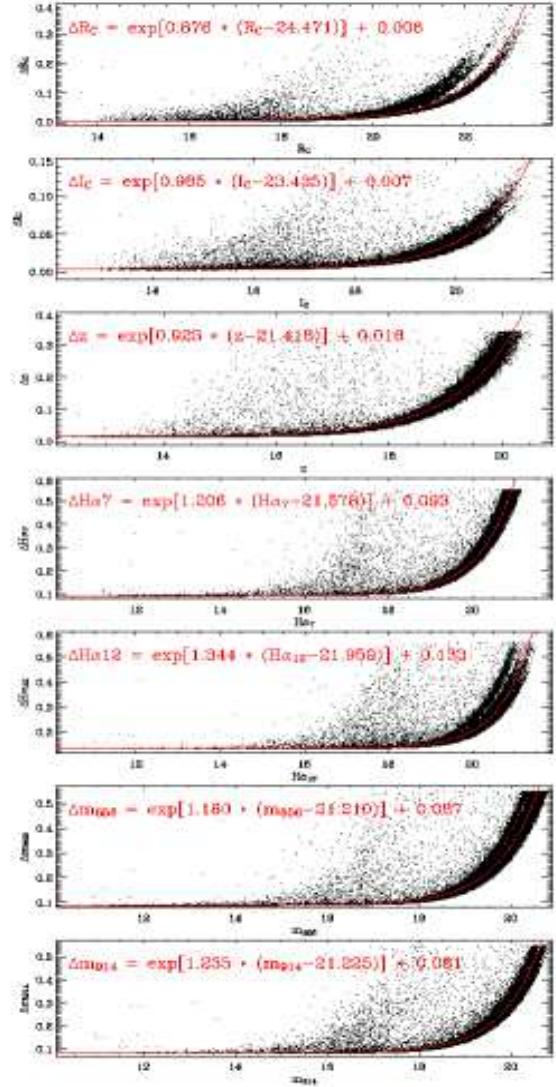}}
\caption{Photometric errors versus magnitudes and relative exponential fits 
for all the point-like sources detected in Cha~II and for all the available 
filters. The ``double sequence'', clearly visible in $R$ and $H\alpha_{12}$ bands, 
is due to different seeing conditions in the individual exposures.}
\label{err}
\end{figure}
%--------------------------------------------------------------------------------

\subsubsection{PSF fitting procedure}

PSF fitting photometry has been performed by 
using the IRAF/daophot package \citep{Ste87}. Since we are interested in the 
faintest stellar objects, the threshold level was defined in order to select 
all the sources having a signal-to-noise ratio ($\sigma$) greater than 3. 
This choice may lead to the extraction of many spurious detections, however 
\emph{daophot} allows us to discriminate between stars, extended sources (galaxies), 
saturated objects and other spurious detections. The PSF fitting procedure was 
re-iterated  twice for each Cha\,II field; given the moderate crowding 
of these fields, this choice allowed us to detect the faintest neighbours 
to bright stars while keeping the number of spurious detections relatively small. 
The typical residual of the PSF fitting is less than 2\% of the peak brightness.  
A single catalogue comprising all the stellar sources detected  in all 
the available photometric bands was finally produced. 
In Fig.~\ref{err} the internal photometric errors of all the detected point-like sources are plotted against the 
magnitude for all the available filters; the relative exponential fits are over-plotted. 
Tab.~\ref{tab:psf} summarises the number of point-like sources detected 
in the surveyed area in each filter and the limiting magnitudes achieved at 
the 10$\sigma$, 5$\sigma$ and 3$\sigma$ levels, respectively.
%These magnitude limits should enable us to detect faint objects down to the 
%low-mass end of the BD domain in the range age 2-5\,Myr, 
%at the distance of Cha~II (178~pc, \citet{Whi97}).

\subsubsection{Completeness}

\begin{figure} 
\resizebox{\hsize}{!}{\includegraphics[width=15cm,height=10cm]{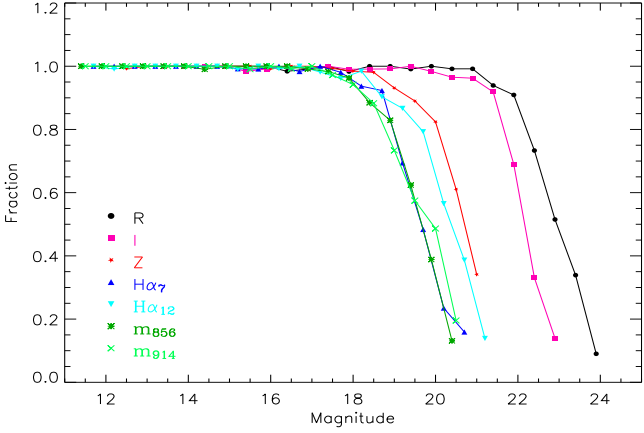}}
\caption{Completeness plots for extraction of artificial stars from the 
ChaII\_2 mosaic (see Tab.~\ref{tab:obs}) for all the photometric bands 
used in this work (see text).}
\label{compli}
\end{figure}

The completeness of our catalogues was estimated in the standard way by 
inserting artificial stars into the images and recovering them using the same
extraction parameters as for the real objects; the fraction of the recovered 
artificial objects provides a measure of the completeness. 
We used the IRAF/addstar package to perform the exercise. We inserted 3000 
artificial sources; this number should not alter significantly the crowding 
statistics in the images. The profile for the artificial sources was generated 
by using the same PSF model used for the source extraction; the position of the
artificial objects are randomly distributed over the entire area of the mosaic 
and their magnitudes range uniformly between the detection and the saturation 
limits in each band. 
As an example we show the results for the ChaII\_2 field (see Tab.~\ref{tab:obs}). 
Fig.~\ref{compli} shows the fraction of recovered artificial objects as a function of
magnitude for each filter. The corresponding magnitude limits at 95\% completeness 
level (C=95\%) are reported in Tab.~\ref{tab:psf}.

%----------------------------------completeness-------------------

%--------------------------------------------------------------------------------

%---------------------------------- Table -------------------
\begin{table}
\caption[ ]{\label{tab:psf} 
Number of stellar sources (N$^*$) detected in each filter and relative 
limiting magnitudes at the 10$\sigma$, 5$\sigma$ and 3$\sigma$ level. 
The last column report the magnitude limit at 95\% completeness level.}
\begin{center}
\scriptsize
\begin{tabular}{cccccc}  
\hline
Filter &  N$^*$ & Mag 10$\sigma$ & Mag 5$\sigma$ & Mag 3$\sigma$ & Mag (C=95\%)\\
\noalign{\medskip}
\hline
\noalign{\medskip}
Rc/162        & 141400 & 21.8  & 22.6  & 23.2 & 21.3 \\
Ic/Iwp        & 141000 & 21.0  & 21.7  & 22.3 & 21.0 \\
z+/61         & 79000  & 18.7  & 19.6  & 20.2 & 18.8 \\
H$\alpha$/7   & 72500  & 17.5  & 19.8  & 20.4 & 18.0 \\
H$\alpha$/12  & 84200  & 17.8  & 19.9  & 20.8 & 18.4 \\
856-nm        & 76700  & 18.6  & 19.4  & 20.0 & 18.0 \\
914-nm        & 80400  & 18.9  & 19.5  & 20.1 & 18.0 \\
\noalign{\medskip}
\hline
\end{tabular}
\end{center}
\end{table}
%--------------------------------------------------------------------------------

%\subsubsection{Comparison with SExtractor}

%For each field, we also performed aperture photometry using the 
%SExtractor tool \citep{Ber96}.
%Although more time-consuming, we found that the PSF fitting method 
%yields lower internal errors and, hence, a better signal-to-noise ratio 
%(by about 15\%) than SExtractor, in agreement with previous investigations 
%\citep{Lar05}. 
%We also estimated a systematic offset of 0.04 mag in both the SExtractor 
%\emph{aperture} and \emph{adaptive aperture} magnitudes with respect to the 
%PSF magnitudes, although the \emph{aperture} magnitudes show a larger scatter.

\section{Tools for candidates selection}
\label{analysis}

Our primary criterion for the selection of low-mass PMS star and BD candidates 
% from our survey in Cha~II is 
from the extracted catalogue was based on the comparison of the object location 
in CMDs with theoretical isochrones. 
In addition, our WFI data in the 865-nm and 914-nm intermediate-bands allowed us 
to obtain a first estimate on the effective temperature of the candidates, whereas 
the measurements in H$\alpha$ provided us with a diagnostics for possible 
H$\alpha$ emission. Our selection method exploits the tools which are now 
described.

\subsection{Theoretical Isochrones}
\label{WFIisoc} 

Theoretical isochrones for low-mass stars and BDs down to 0.001~M$_{\odot}$ 
are provided by \citet{Bar98} and \citet{Cha00} in the Cousins photometric system \citep{Bes90}. 
Since the WFI filters transmission curves are somewhat different from the 
original Cousins ones, in particular for the $I$-band, it is crucial to transform 
the colours and magnitudes into the appropriate photometric system. 
We thus transformed the isochrones by \citet{Bar98} and \citet{Cha00} into 
the WFI-Cousins system as described in Appendix~\ref{isocs}. 
In this way, we can use the isochrones to define the PMS locus in the CMDs 
(e.g. Fig.~\ref{fig:opt_CMDs}) in a photometrically consistent way.

\begin{figure} 
\resizebox{\hsize}{!}{\includegraphics[width=10cm,height=8cm]{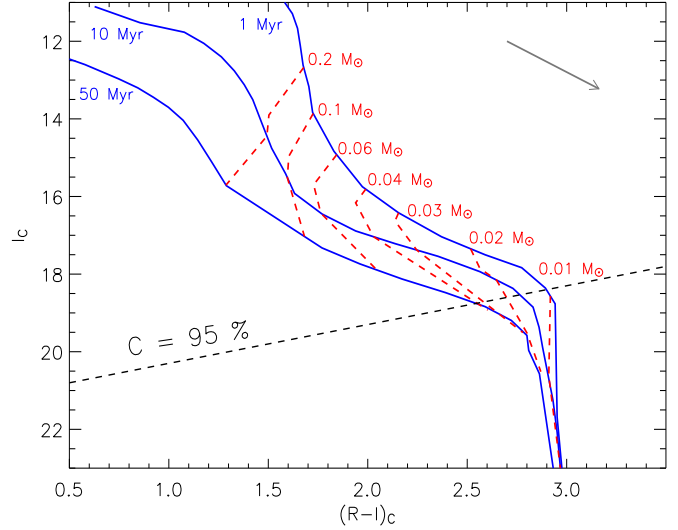}}
\caption{Theoretical $R_C$ vs. ($R$-$I$)$_C$ diagram. The isochrones (continuous curves) and 
PMS tracks (dashed curves), shifted to the distance modulus of Cha~II (6.25~mag, \citet{Whi97}) 
are in the WFI-Cousins photometric system. The dashed straight line represents the 95\% 
completeness limit. The A$_{\rm V}$=2~mag reddening vector is shown.}
\label{compli_mass}
\end{figure}

In Fig.~\ref{compli_mass} the theoretical $R_C$ vs. ($R$-$I$)$_C$ colour-magnitude 
diagram is shown. The 95\% completeness limit was determined from the values
reported in Tab.~\ref{tab:psf}. For A$_V$=0 we are complete down to 0.02~M$_{\odot}$ 
at the 95\% level for objects younger than 10 Myr. 
For an average extinction A$_V\approx$2~mag for Cha\,II \citep{Cam99} a 95\% 
completeness limit corresponds to 0.03~M$_{\odot}$.

%-------------completeness mass-------------------------------------------
%-----------------------------------------------------------------------------

\subsection{The ($m_{856}-m_{914}$) index vs. temperature relation} 
\label{Teff}

Late-type dwarfs are characterised by strong molecular absorption bands 
essentially due to metallic oxides (TiO and VO). The ESO-WFI medium-band 
filter centered at 856\,nm  covers important TiO absorption features 
that deepen with decreasing temperature, while the medium-band filter 
centered at 914 nm lies in a wavelength range which is relatively featureless 
in late-type objects (see Fig.~\ref{fig:filter} lower panel in Appendix~\ref{isocs}). 
Thus, the ($m_{856}-m_{914}$) colour index is sensitive to the effective 
temperature for very cool objects (2000~K$\lesssim$T$_{\rm eff}\lesssim$3800~K), 
where TiO dominates the opacity \citep{All90,Bur93,ONe98}. In the absence
of interstellar extinction, this index can provide a first reliable guess 
of the temperature of the cool candidates.

Using the filter transmission curves for the corresponding intermediate-band 
filters and the synthetic low-resolution StarDusty spectra for low-mass 
stars and BDs by \citet{All00} we derived the relationship between the 
($m_{856}-m_{914}$) index and the effective temperature, applying the methods 
described in Appendix~\ref{isocs}. 
The synthetic ($m_{856}-m_{914}$) index, derived by integrating the 
StarDusty spectra under the filter transmissions and CCD 
quantum efficiency curves, is in the instrumental WFI system. 
As such, it was transformed into the standard AB photometric system 
by applying the zero points derived from the spectrophotometric standard 
stars Hiltner~600, LTT~4364 and Eg~274, observed with WFI in the 856-nm 
and 914-nm bands (see Sec.~\ref{photcal}). 
% From their low-resolution spectra 
From the standard fluxes \citep{Ham92} integrated under the WFI 
passbands likewise the StarDusty spectra, we found an average zero-point 
correction for the ($m_{856}-m_{914}$) colour of $-$0.13~mag, which allowed us 
to obtain the calibrated synthetic (m$_{856}-m_{914}$) index as:

\begin{equation}
(m_{856}-m_{914})^{synt} _{calib}=(m_{856}-m_{914})^{synt} _{instr}-0.13
\end{equation}

\begin{figure} 
\resizebox{\hsize}{!}{\includegraphics[width=3cm,height=2cm]{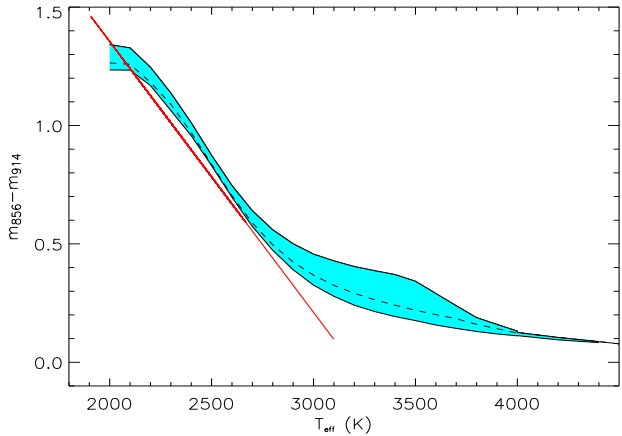}}
\caption{($m_{856}-m_{914}$) index versus effective temperature relation 
derived from the corresponding WFI intermediate-band filters and the 
synthetic low-resolution StarDusty spectra for low-mass stars and BDs 
by \citet{All00}. The two solid curves show the relation for 
$\log g$=5.0 (lower boundary) and $\log g$=3.5 (upper boundary); the 
dashed curve represents the T$_{\rm eff}$-colour relation for $\log g$=4.0. 
The straight line is the empirical relationship derived by \citet{Lop04}.}
\label{TeffInt}
\end{figure}

In Fig.~\ref{TeffInt} we show the resulting ($m_{856}-m_{914}$) 
index versus temperature relation. This relation depends on the 
stellar gravity, as indicated by the shaded area in Fig.~\ref{TeffInt}, 
which encompasses all the calibrations for models in the range 
3.5$\leq \log g \leq$5. 
We also derived this relationship using other stellar atmosphere 
models (e.g. NextGen, COND, etc.); the StarDusty models were chosen 
because they better match the empirical relationship between the 
($m_{856}-m_{914}$) index and the effective temperature by \citet{Lop04} 
(straight line in Fig.~\ref{TeffInt}). 
Furthermore, \citet{All01} found that silicate dust grains can form 
abundantly in the outer atmospheric layers of the latest M dwarfs 
and BDs. 

In the case of high interstellar extinction (A$_V \gtrsim$4~mag), 
the temperature derived from this relation may be 
underestimated because the ($m_{856}-m_{914}$) index becomes larger 
than in the absence of extinction. Thus, without any reliable 
evaluation of A$_{\rm V}$, we can use these temperature estimates only 
for candidate selection purposes (see Sec.~\ref{sel_cand}).

%---------------------------------- Teff.vs.856-914-------------------
%-------------------------------------------------------------- 

\subsection{H$\alpha$ Equivalent Width calibration}
\label{cal_ha_ew}

%---------------------------------- EW_Ha.vs.Ha12-Ha7-------------------
\begin{figure} 
\centering
\includegraphics[width=7cm,height=7cm]{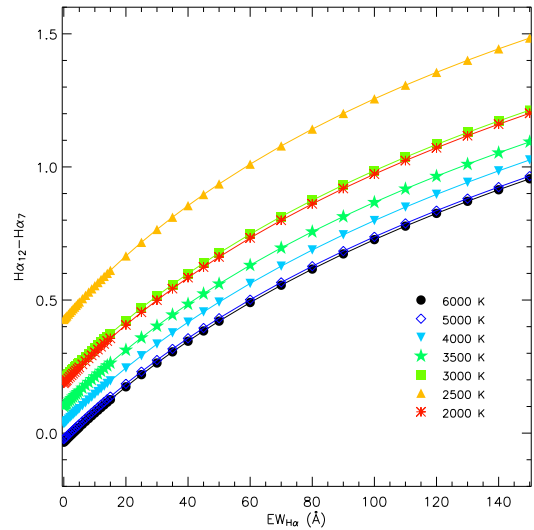}
\caption{Calibration relation between the ($H\alpha_{12}-H\alpha_7$) index 
and the H$\alpha$ equivalent width derived from the corresponding 
WFI intermediate-band filters and the synthetic low-resolution 
StarDusty spectra for low-mass stars and BDs by \citet{All00}.
}
\label{calHa}
\end{figure}
%-------------------------------------------------------------- 

The WFI $H\alpha_{7}$ filter covers the H$\alpha$ line, while the 
$H\alpha_{12}$ filter lies in a wavelength range which is affected 
by neither the H$\alpha$ wings nor strong photospheric lines 
(see Fig.~\ref{fig:filter} mid panel in Appendix~\ref{isocs}). 
The ($Ha_{12}-Ha_{7}$) index is thus sensitive to the intensity of the
H$\alpha$ line. 
We used the StarDusty spectra by \citet{All00} to define a calibration 
relation between this index and the H$\alpha$ equivalent width (EW$_{\rm H\alpha}$). 
In order to simulate stellar spectra with different values of EW$_{\rm H\alpha}$, 
we added to the synthetic spectra a Gaussian emission profile centered 
at H$\alpha$ (656.2 nm) and having a FWHM of 3 \AA. The latter value 
was fixed by considering the H$\alpha$ line profiles of a sample of 75 PMS 
stars presented by \citet{Fer95}; we found that the typical FWHM of the 
H$\alpha$ emission lines of these objects is around  3~\AA, though it 
can be higher for classical T~Tauri stars (5-6~\AA). However, from 
tests made by using profiles with different FWHM, we have verified that 
the FWHM value has no significant effect on the 
($H\alpha_{12}-H\alpha_7$) vs. EW$_{\rm H\alpha}$ 
calibration.

By applying the method described in Appendix~\ref{isocs} and 
correcting the synthetic ($H\alpha_{12}-H\alpha_7$) colour by the mean 
($H\alpha_{12}-H\alpha_7$) colour of the spectrophotometric standard 
stars used to transform the intermediate-band photometry 
into the AB system (see Sec.~\ref{photcal}), we obtained the calibration 
relation ($H\alpha_{12}-H\alpha_7$) vs. EW$_{H\alpha}$ displayed in 
Fig.~\ref{calHa}. 
This relation does not depend on the FWHM of the Gaussian emission core 
but is strongly dependent on the effective temperature for objects cooler 
than about 4500~K. As expected from the contribution of the H$\alpha$ 
absorption wings (see Fig.~\ref{fig:filter}, mid panel), for a fixed 
EW$_{\rm H\alpha}$ the ($H\alpha_{12}-H\alpha_7$) index is higher for 
cooler objects; between 3000 and 2000~K this trend is inverted because 
the $H\alpha_{12}$ filter covers strong molecular absoption bands that 
heavily affect the continuum of very cool stars.

%---------------------------------- Ha12-Ha7 synth-obs-------------------
\begin{figure} 
\centering
\includegraphics[width=7cm,height=7cm]{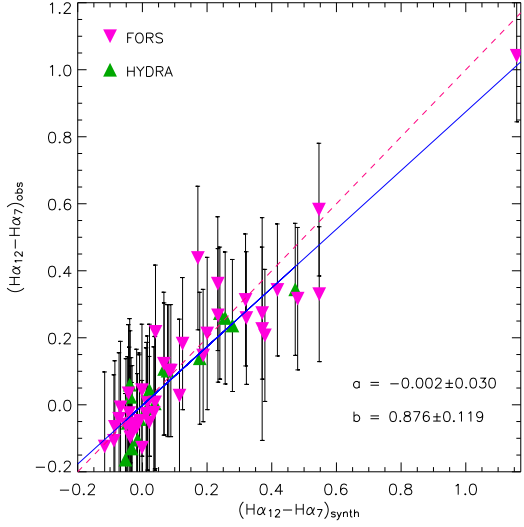}
\caption{($H\alpha_{12}-H\alpha_{7}$) observed colours for PMS stars in L1616 
as a function of the synthetic ($H\alpha_{12}-H\alpha_{7}$) colours (see text). 
The continuous line represents the best fit of the data; 
the dashed line has offset a=0 and slope b=1.}
\label{synt_obs}
\end{figure}
%-------------------------------------------------------------- 

%---------------------------------- EW_Ha   phot vs spec-------------------
\begin{figure} 
\centering
\includegraphics[width=9cm,height=7cm]{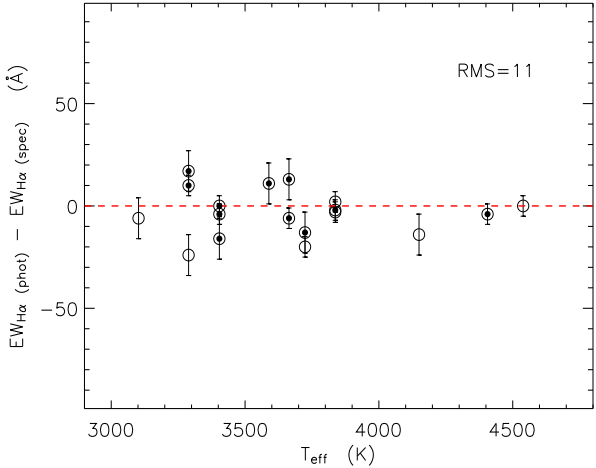}
\caption{Comparison between the H$\alpha$ equivalent width estimated 
in this work from the ($H\alpha_{12}-H\alpha_{7}$) photometric index 
and that measured by \citet{Hug92} and \citet{Bar04} for confirmed 
PMS stars in Cha~II (open circles). The smaller black dots mark the 
objects recovered by our selection criteria (see Sec.~\ref{sel_cand}).}
\label{check_ha}
\end{figure}
%-------------------------------------------------------------- 

The relationship between the H$\alpha$ index and EW$_{\rm H\alpha}$ derived 
above was checked by using unpublished WFI H$\alpha$ photometry and 
spectroscopy of already confirmed PMS stars in the L1616 cometary cloud 
obtained with HYDRA@WIYN3.5m \citep{Alc04} and FORS@VLT (Alcal\'a et al., 
in preparation). 
The spectra of the PMS stars in L1616 were integrated under the transmission 
curves of the WFI H$\alpha$ filters to derive their ``synthetic'' 
($H\alpha_{12}-H\alpha_{7}$) index. Such index was then transformed into 
the AB system by using the mean H$\alpha$ colour index of the corresponding
spectrophotometric standard stars observed in the L1616 run. 
The WFI H$\alpha$ photometry of these PMS stars was then used to compare
the observed ($H\alpha_{12}-H\alpha_{7}$) index with the one derived from
the spectra. Fig.~\ref{synt_obs} shows the results. The best fitting line 
(continuous line in Fig.~\ref{synt_obs}) has a slope close to $1$ and an offset 
close to zero; this implies that no significant correction must be applied to 
the ($H\alpha_{12}-H\alpha_{7}$) vs. EW$_{\rm H\alpha}$ relation deduced from 
the synthetic spectra. Based on the errors of the observed 
($H\alpha_{12}-H\alpha_{7}$) index we conclude that {\em bona-fide} emission
line objects will have an H$\alpha$ index greater than 0.1, 
corresponding to EW$_{\rm H\alpha} \approx 10\ \AA$ for an object with 
T$_{\rm eff} \approx$ 4000~K.

The H$\alpha$ index versus EW$_{\rm H\alpha}$ relation was also checked using
data from the literature for the Cha~II PMS stars. In Fig.~\ref{check_ha} 
the H$\alpha$ equivalent width derived from spectroscopic data \citep{Hug92}, 
EW$_{\rm H\alpha}\ (\rm spec)$, is compared with the one derived from the H$\alpha$ 
photometric index, EW$_{\rm H\alpha}\ (\rm phot)$. 
In order to estimate the latter, the effective temperatures reported 
by \citet{Hug92} were used, while for the PMS star C~41 the temperature 
reported by \citet{Bar04} was used. From Fig.~\ref{check_ha}, the EW$_{\rm H\alpha}$ 
values derived from the H$\alpha$ photometric index are in good agreement 
with those derived from spectroscopic data. The computed RMS suggests 
that the relationship shown in Fig.~\ref{calHa} can be used to estimate 
the EW$_{\rm H\alpha}$ from the WFI H$\alpha$ photometry for objects with 
EW$_{\rm H\alpha} \gtrsim 10\ \AA$ and EW$_{\rm H\alpha} \gtrsim 30\ \AA$, 
at the 1$\sigma$ and 3$\sigma$ levels respectively, provided we have an 
estimate of the stellar temperature.

\section{Selection of the candidates}
\label{sel_cand}

For the selection of low-mass PMS star and young BD candidates we exploited 
both the optical data from our survey and the $JHK$ photometry available 
from the 2MASS catalogue \citep{Cut03}, using the tools described in the 
previous sections. A matching radius of 0.5\arcsec was used to merge the
WFI and the 2MASS catalogues; this value was set by taking into account 
the astrometric accuracy of both catalogues.   

Low-mass PMS star and BD candidates in SFRs can be identified in optical CMDs. 
In such diagrams the colour increases rapidly for late spectral types and hence, 
the contamination from foreground stars is expected to decrease. 

%---------------------------------- CMDs-------------------
\begin{figure} 
\resizebox{\hsize}{!}{\includegraphics[width=2.5cm,height=2cm]{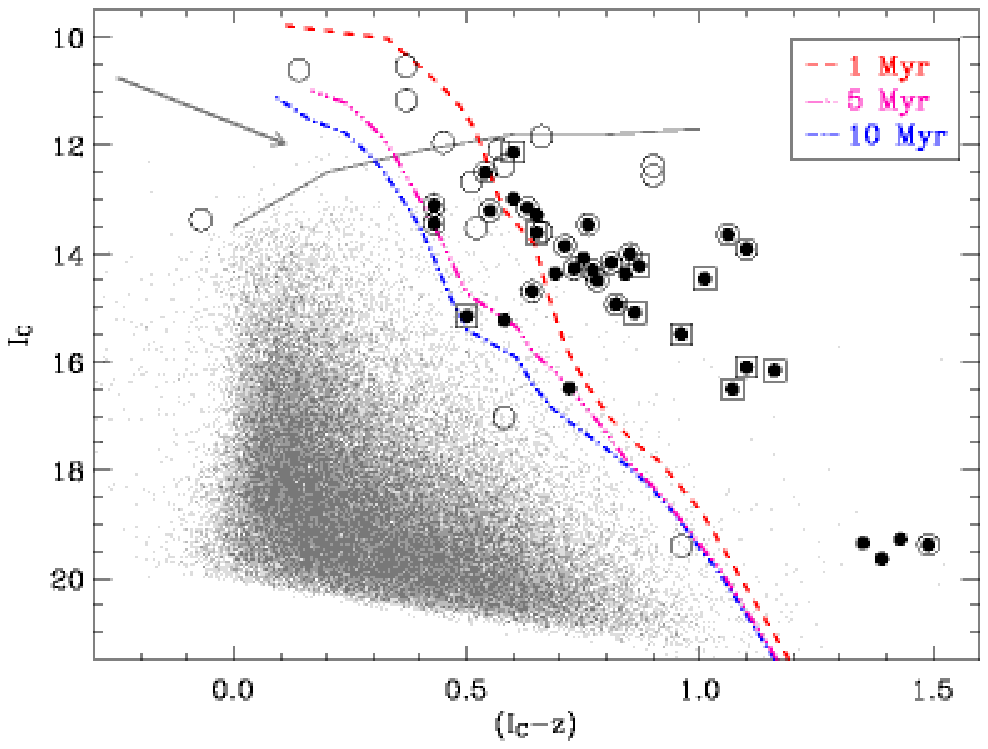}}
\resizebox{\hsize}{!}{\includegraphics[width=2.5cm,height=2cm]{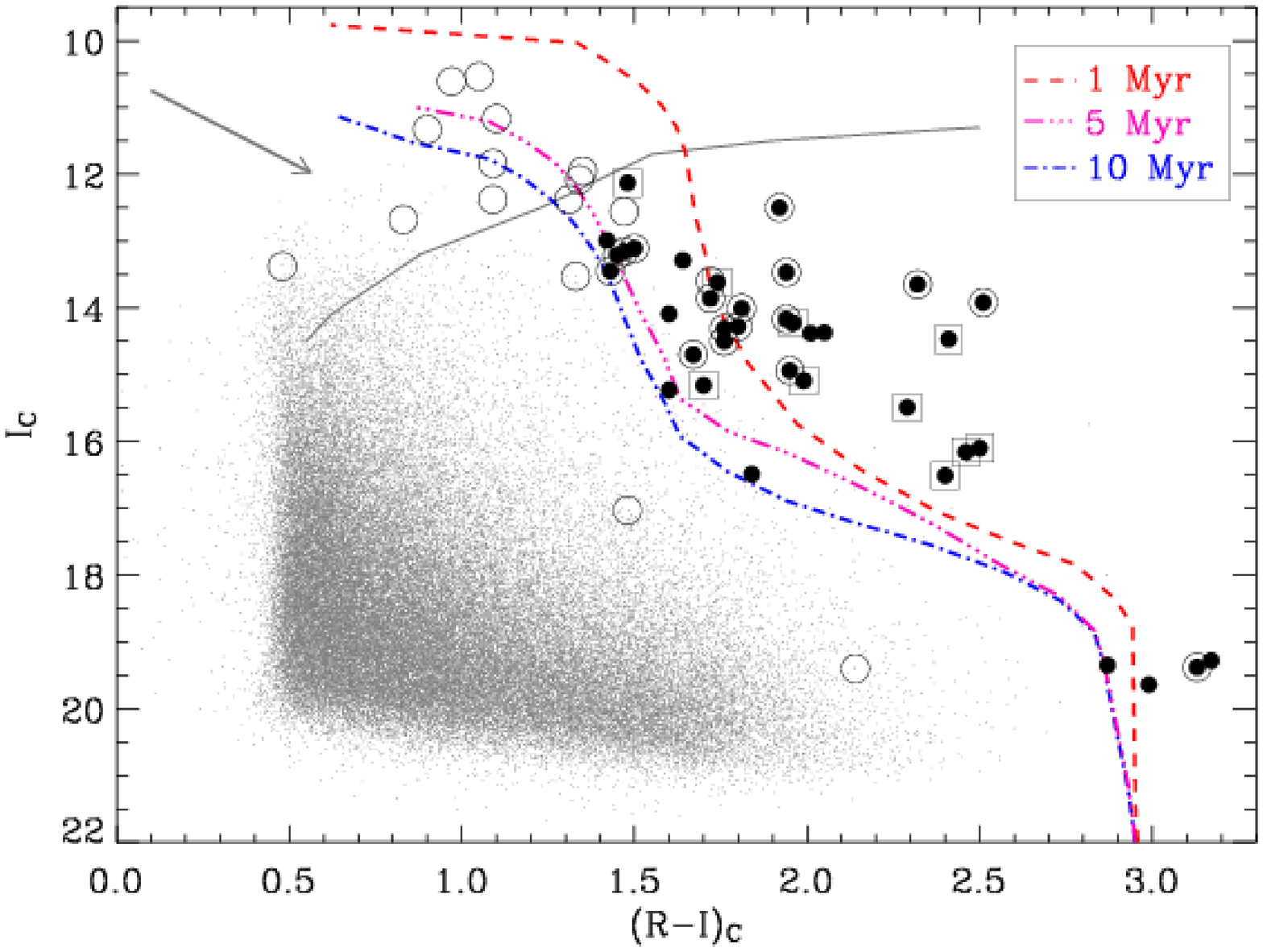}}
\caption{$I_C$ vs. ($I_C-z$) and $I_C$ vs. ($R-I$)$_C$ diagrams for the 
point-like objects in our survey (small gray dots). The lines represent the theoretical 
isochrones derived as explained in Sec.~\ref{WFIisoc}, shifted to the 
distance modulus of Cha~II (6.25 mag, \citet{Whi97}). 
The continuous line in each diagram represents the saturation limit. 
The big black dots represent the 37 sources selected from our 
criteria, while the open circles and squares represent the known 
PMS stars and candidates selected from previous surveys respectively. 
The magnitudes reported by \citet{Hug92} were used to plot the objects
above and/or close to our saturation limits in any of the three 
broad-bands. Note that basically all the known PMS stars and candidates 
are recovered in our selection. The A$_{\rm V}$=2\,mag reddening vector is 
shown in both diagrams.}
\label{fig:opt_CMDs}
\end{figure}
%-------------------------------------------------------------
A first sample of candidates was then selected based on the 
$I_C$ vs. ($I_C-z$) and $I_C$ vs. ($R-I$)$_C$ CMDs. All the point-like 
sources detected in Cha~II above the 3$\sigma$ level were placed in 
these CMDs (Fig.~\ref{fig:opt_CMDs}) and their position was compared 
with the theoretical isochrones. The isochrones transformed into the 
WFI-Cousins system, as described in Sec.~\ref{WFIisoc}, were scaled 
to the Cha~II distance of 178~pc \citep{Whi97} in both CMDs. 
We then selected all the objects falling above the 10~Myr isochrone 
in both CMDs (Fig.~\ref{fig:opt_CMDs}). This criterion takes into 
account the age spread of the previously known PMS stars in Cha~II 
\citep[0.1-10~Myr,][]{Hug92} and the PMS locus that they define on 
the CMDs, as well as the uncertainties in interstellar extinction. 
In this way we selected 114 objects of which 17 were previously known 
Cha~II members.
 
%---------------------------------- near IR diag.-------------------
\begin{figure} 
\resizebox{\hsize}{!}{\includegraphics[width=3cm,height=2.5cm]{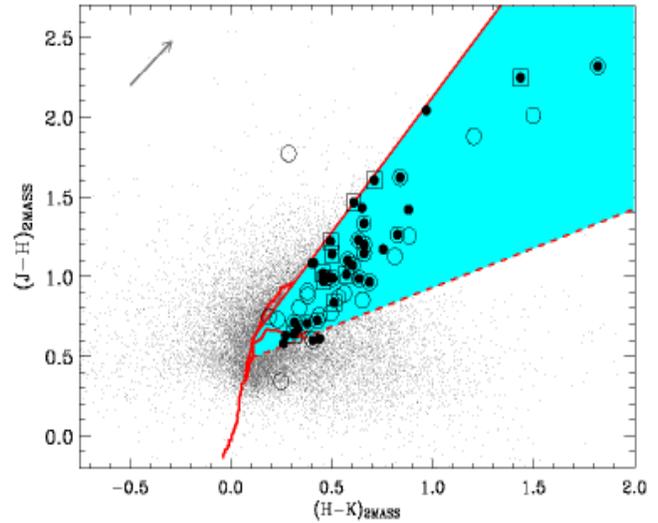}}
\caption{($J-H$) vs. ($H-K$) diagram for the sources detected in Cha~II (small gray dots). 
The solid curve shows the relation between these indexes for main sequence 
stars (lower branch) and giants (upper branch). The dashed line indicates 
the classical T~Tauri stars locus by \citet{Mey97}. The continuous line 
is dividing line between giants and dwarfs. 
The shaded area indicates the locus presumably uncontaminated by giants.
The previously known PMS stars and candidates are indicated with open 
circles and squares respectively. The big black dots represent the objects
selected in this work. The A$_{\rm V}$=2~mag reddening vector is shown. The
two PMS stars which fall outside the selection area are the heavily 
veiled star Sz~47 (below the shaded area) and the binary Sz~60E 
(above the shaded area).}
\label{fig:JHK_diag}
\end{figure}
%-------------------------------------------------------------- 

However, the 97 new candidates from the selection above may still be 
contaminated by background giants and highly-reddened objects. 
We therefore investigated the properties of the selected sample in 
the $JHK$ bands using their 2MASS magnitudes. It has been found by 
different authors that the intrinsic ($J-H$) and ($H-K$) colours of 
young late-M objects are dwarf-like with, in some cases, an ($H-K$) 
excess, mainly arising from a circumstellar disk or an in-falling 
envelope \citep{Luh99a, Lee05, Mey97}. 
Furthermore, \citet{Mey97} found that classical T~Tauri stars, with 
prominent IR excesses due to circumstellar accretion disks, exhibit 
a narrow range of colours in the ($J-H$) vs. ($H-K$) diagram (dashed 
line in Fig.~\ref{fig:JHK_diag}). 
By using the ($J-H$) vs. ($H-K$) diagram we then performed a secondary 
selection on the sample coming from the optical CMDs, following similar
criteria as in \citet{Lee05}. The sub-sample of all the objects falling 
below the dividing line between giants and dwarfs 
(c.f. Fig.~\ref{fig:JHK_diag}) should be less contaminated by background 
giant stars. In this way we end up with a sub-sample of 84 objects, 
which still include the 17 previously known members.

The sub-sample passing the secondary selection may still 
contain foreground and highly-reddened background objects. 
We thus performed a third-level selection as follows.
We assumed that the objects passing the primary and secondary
selections above are all at the distance of Cha~II and estimated  
their temperature and luminosity as described in Sec.~\ref{sec:par}. 
Although these are first guesses of the actual stellar 
temperatures and luminosities, these values already provide 
good estimates for candidate selection purposes. 
We then constructed the HR diagram of our candidates sample and 
selected all the objects falling between the birth-line and the 
20\,Myr isochrone, using the models by \citet{Bar98} and \citet{Cha00}. 
Although the age distribution of the confirmed members of Cha~II 
ranges between 0.1 and 10 Myr \citep{Hug92}, the more relaxed 
criterion of a 20 Myr cut-off for the candidates with a high 
membership probability was set by taking into account the 
uncertainties on the distance to the Cha~II cloud 
\citep[$\Delta$d=18~pc,][]{Whi97} and the interstellar 
extinction. 
The evolutionary tracks of low-mass PMS stars run almost parallel to the luminosity axis of the H-R diagram. 
Thus, though mass estimates are not greatly affected by interstellar extinction, the age dispersion may be 
considerably increased if the adopted extinction for each candidate is not well determined.
%Indeed, while mass estimates are not greatly affected by the 
%interstellar extinction, being the evolutionary tracks of low-mass 
%PMS stars  almost parallel to the luminosity axis in the HR diagram, 
%the age dispersion may be considerably increased if the adopted 
%extinction for each candidate is not well determined.
The mean uncertainty on our visual extinction determinations is $\sim$1.5~mag (see Sec.~\ref{seds}); 
this would correspond to a luminosity uncertainty of $\sim$16\% and, hence, and age uncertainty of 
$\sim$2~Myr for an object with mass $\sim$0.5M$_{\odot}$ and age 2-3~Myr, 
i.e. the typical values in Cha~II (Alcal\'a et al., in preparation).
Thus, the criterion of the 20\,Myr isochrone ensures that the 
selected objects are, to a first approximation, consistent with 
Cha~II membership.

\begin{figure*} 
\resizebox{\hsize}{!}{\includegraphics[width=2.5cm,height=2cm]{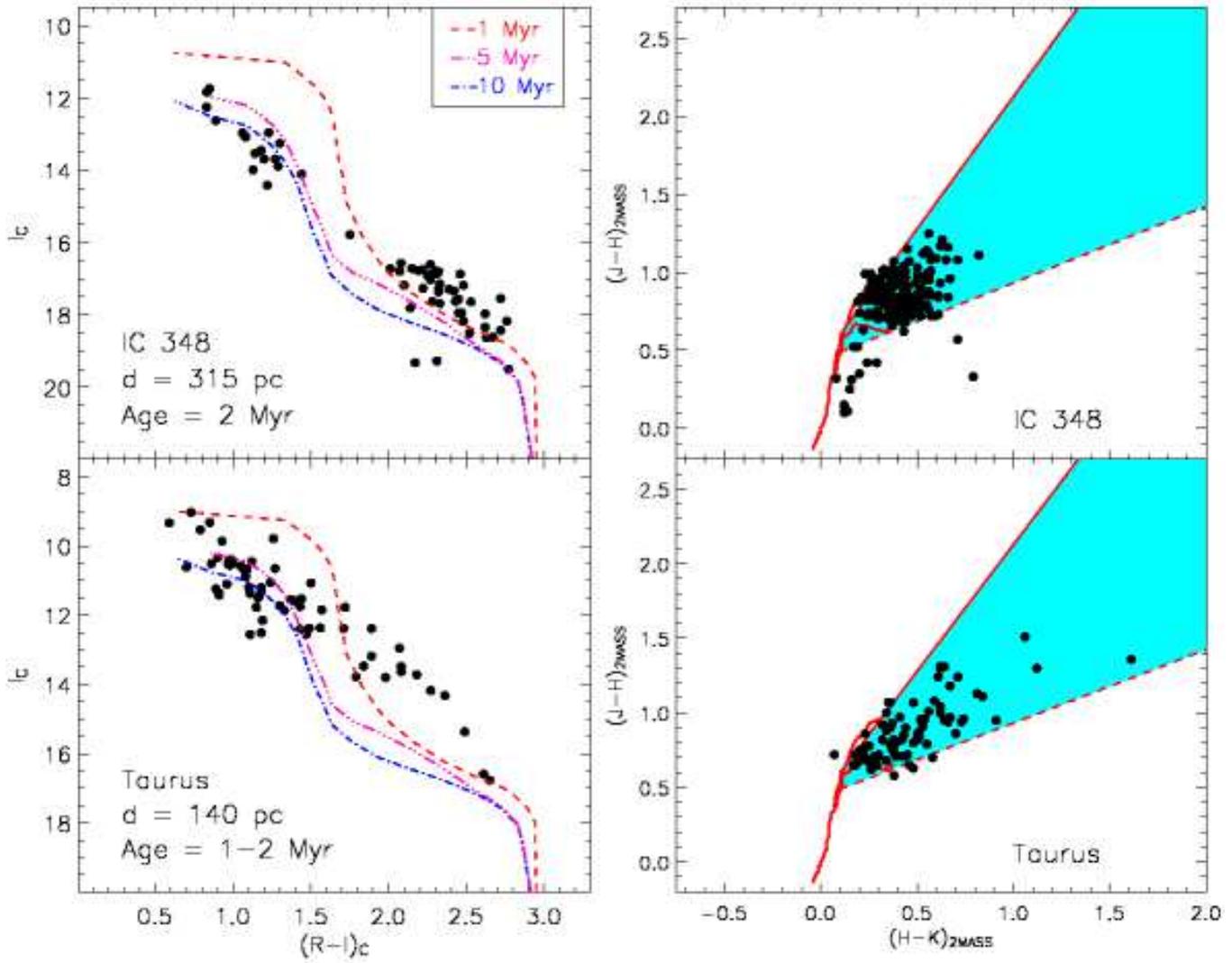}}
\caption{$I_C$ vs. ($R-I$)$_C$ and ($J-H$) vs. ($H-K$) diagrams for the PMS populations 
in Taurus and IC348. The isochrones, shifted by the corresponding distance
modulus, are as in Fig.~\ref{fig:opt_CMDs}. The selection limits in the
($J-H$) vs. ($H-K$) diagrams are as in Fig.~\ref{fig:JHK_diag}.
}
\label{test_criteria}
\end{figure*}

The theoretical isochrones depend on the physics involved in the 
models. As such they may be rather uncertain, in particular in the very 
low-mass domain. We have thus tested the reliability of our selection 
criteria using as test-bench the publicly available samples of 
confirmed PMS stars and BDs in Taurus and IC348 reported by \citet{Bri02} 
and \citet{Luh03}, respectively. As can be seen in Fig.~\ref{test_criteria} 
(available only in electronic form), the selection criteria recover
the vast majority  of the previously known PMS stars and BDs in 
these regions, providing a good check that they work well on
selecting these type of objects.

\section{Results of the survey}
\label{sec:par}

The already confirmed Cha~II population members consist 
of some 36 objects, comprised of 33 PMS stars \citep{Hug92, Alc00, You05},  
one  BD \citep{Alc06}, and two planetary-mass objects 
\citep{All06, Jay06}.

The screening of our data lead, on the other hand, to the selection 
of 37 interesting objects. Of these, 17 were already known (16 PMS stars 
plus the BD) and 10 are candidates with high probability of membership 
based on their selection in previous surveys \citep{Vuo01, Per03, You05, All06}. 
The remaining 10 selected objects are completely new candidates. 
The other 13 previously known PMS stars are saturated and/or close 
to the saturation limit in at least one of our broad-band $RIz$ 
images and hence, we could not apply our selection criteria. 
However, using their magnitudes and colours from the 
literature \citep{Hug92, Alc95}, these objects would fall within 
our selection; this means that, adding these to the 17 previously 
known non-saturated objects, our criteria recover 30 of the 36
previously known Cha\,II members.

There are a few exceptions in which the above selection criteria fail:
the heavily veiled T~Tauri stars Sz~47 \citep{Hug92} and C~41 \citep{Bar04}, 
which exhibit strong UV excess that affects their colours, and 
the Class-I source IRAS~12500-7658 \citep{You05}, which is a deeply 
embedded object. These three sources are sub-luminous in both CMDs.
In addition, the other Class-I source, ISO-CHA\,II\,28, is not detected
in any of our images. Although this type of objects would escape our 
selection, they are expected to be rare in Cha~II (\citet{You05}; 
Alcal\'a et al., in preparation). The two planetary-mass candidates 
reported by \citet{All06} also escaped selection because they were 
barely detected only in our I-band images. 
The Herbig-Haro object HH~54 \citep{Gia06} and the Class-0 source 
BHR~86 \citep{Gar02} 
are also associated with Cha~II. HH~54 was detected only in our $R$-band 
and H$\alpha$ images, while BHR~86 was not detected in any of the optical 
bands. These objects will be discussed in detail in a forthcoming paper 
(Alcal\'a et al., in preparation).

In conclusion, our selection criteria recover the majority 
(about 80\%) of the confirmed and candidate members of 
Cha~II reported in previous surveys. For the sake of clarity, 
in all the diagrams we use dots to represent the 37 objects 
selected with our criteria, open circles to represent the 
already confirmed members and open squares to represent the 
10 candidates selected in previous surveys. The 10 new candidates 
appear simply as dots in all the diagrams. 

In Tab.~\ref{tab:phot} we report all the objects with at least one 
photometric measurement in the optical bands\footnote{Table~\ref{tab:phot} 
is published only in electronic form.}. 
The table contains the confirmed members of Cha~II, the candidates 
selected by previous surveys and the new candidates selected in this 
work, as well as other sources identified as possible candidates in 
previous studies, but which were rejected by our selection criteria. 
The latter sample includes objects like the candidates reported by 
\citet{Lop05}\footnote{The BD candidates selected by \citet{Lop05} do 
not comply with our selection criteria, probably because the colour 
effects on the isochrones due the WFI filters was not taken into 
account.} and some of the sources identified by \citet{Per03}. 
Comments on some individual objects are given in Appendix~\ref{comm}.\\

\noindent
In the following we concentrate on the sample of 20 candidates, 
namely the 10 candidates selected from previous surveys, but 
also recovered from our selection criteria, and the 10 new candidates; 
they are reported in Tab.~\ref{tab:par} and their spatial distribution 
is shown in Fig.~\ref{RaDec}. 
The 10 previously known candidates we recovered in our selection 
have also been proposed as young members of Cha~II by recent studies 
\citep{Vuo01, You05, All06}; this supports their membership to the 
cloud and, at the same time, the reliability of our selection method.
Moreover, several of them possess H$\alpha$ emission (see Sec.~\ref{Ha}).
We remark that these are the candidates with the highest 
membership likelihood but we do not exclude that a few true Cha~II 
members could have escaped some of our selection criteria.
In particular, if they are heavily veiled, deeply embedded or 
components of unresolved binaries. 
If all the 20 candidates will be confirmed by spectroscopy, 
the population of Cha~II will increase to some 56 members. 
Two of the 10 new candidates in our sample (WFI\,J12585611-7630105 
and WFI\,J13005531-7708295) have been spectroscopically confirmed 
to be PMS stars (Alcal\'a et al., in preparation).

\subsection{Possible contaminants} 
\label{contamin} 

One possible source of contamination of our candidates is 
represented by field dwarfs. In order to estimate the number
of possible foreground dwarfs in the field of ChaII we
followed the prescriptions by \citet{Bur04b}.
We used the low-mass luminosity function simulations from 
\citet{Bur04a} and the absolute magnitudes for late M and L field 
dwarfs from \citet{Dah02}. Assuming a limiting magnitude of $I$=20, 
i.e. the $I$ magnitude of our faintest candidates, and a mass 
function $dN/dM \propto M^\alpha$ with 0.5$<\alpha<$1.5, we expect
8-10 foreground dwarfs with 2000$<$T$_{\rm eff}<$3000~K in the
$\sim$2 square degrees area observed in Cha~II. Therefore,
up to about 50\% of the candidates might be foreground
dwarfs unrelated to the SFR.

Another possible source of contamination is represented
by faint galaxies which may have colours similar to those
of PMS objects, in particular of BD candidates. 
From the $K$-band galaxy number counts toward the celestial 
south pole, the expected number of background galaxies with 
$K \lesssim$13~mag, i.e. the $K$ magnitude of our faintest 
candidates, in a $\sim$2 square degrees area is $\sim$20 
\citep{Min98}. For this estimate only diffuse interstellar 
extinction, which is negligible ($\sim$0.02~mag) at the 
$K$-band \citep{Jon81}, is considered. In Cha\,II, the extinction may 
be as high as A$_{\rm V}\approx$6-8~mag, i.e. A$_{\rm K}\approx$1~mag. 
Thus, in order to contaminate our sample, the background 
galaxies should have $K\lesssim$\,12~mag; the predicted galaxy 
number count at this magnitude is less than 2 in the surveyed
area.
In addition, our PSF extraction methods remove the extended 
objects quite efficiently; thus, the only contaminants may 
be eventually the point-like extra-galactic objects, 
mainly QSO's. According to \citet{Pre06} only a handful of 
QSO's are expected to be present in 2 square degrees for 
magnitudes brighter than $R=$20 mag. The number increases 
for fainter magnitudes, but the vast majority of our 
candidates are brighter. Thus, we do not expect that extra-galactic objects represent
a major problem of contamination.

\subsection{H$\alpha$ emission of the candidates}
\label{Ha}

\begin{figure} 
\resizebox{\hsize}{!}{\includegraphics[width=3cm,height=2.5cm]{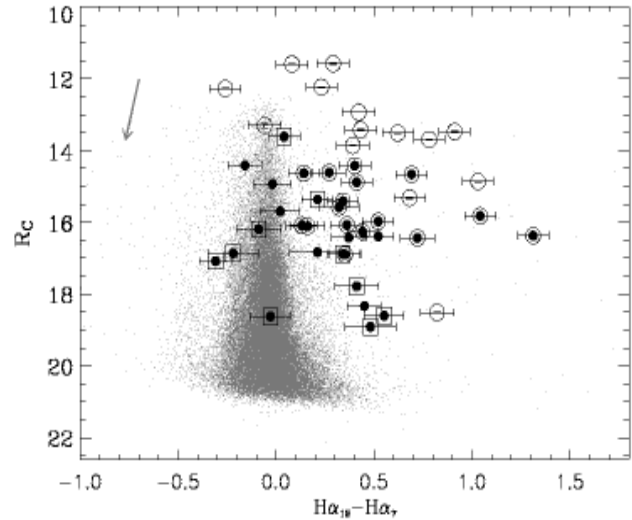}}
\caption{$R_C$ magnitudes versus ($H\alpha_{12}-H\alpha_7$) colours 
for the point-like objects in Cha~II (small gray dots). The previously known PMS stars 
and candidates are indicated by open circles and squares respectively, 
whereas the sources selected in this work are represented with big 
black dots. The A$_{\rm V}$=2~mag reddening vector is shown.}
\label{fig:ha_diag}
\end{figure}

The calibration relation between the ($H\alpha_{12}-H\alpha_{7}$)
index and the EW$_{\rm H\alpha}$ (see Sec.~\ref{cal_ha_ew}) allows 
us to investigate a possible H$\alpha$ emission of the candidates. 
Based on this calibration, a threshold of 
($H\alpha_{12}-H\alpha_7$)=0.1 mag would translate into an equivalenth 
width EW$_{\rm H\alpha}\approx$ 10~\AA~ at the 1$\sigma$ level of detection 
(c.f. Sec.~\ref{cal_ha_ew}). Most of the previously known PMS stars and candidates have an $H\alpha$
index larger than this value (c.f. Fig.~\ref{fig:ha_diag}). 
40\% of the 20 candidates show H$\alpha$ emission 
at the 1$\sigma$ level, but only 2 of the 20 show it at
the 3$\sigma$ level. Considering that the H$\alpha$ line can be 
strongly variable in young objects, we cannot use the index as a 
major diagnostic for the selection of the candidates, but only as a 
consistency check on their possible PMS nature. However, the 1$\sigma$ 
level detections are yet consistent with recent spectroscopic evidence 
of H$\alpha$ emission (Alcal\'a et al., in preparation). 
Moreover, the EW$_{\rm H\alpha}$ range covered by our candidates (10-35~\AA) 
is in line with that found by \citet{Whi03} for the slowly rotating, 
non-accreting stars and BDs in Taurus-Auriga. 
For the remaining candidates the ($H\alpha_{12}-H\alpha_{7}$) 
index does not indicate clear evidence for H$\alpha$ emission. 
For three objects (WFI\,J12533662-7706393, WFI\,J12583675-7704065, 
and WFI\,J13005297-7709478) the level of H$\alpha$ emission could 
not be estimated because they are not detected in our H$\alpha$ 
images (the magnitude limit is $\sim$20.5~mag at the 3$\sigma$ level). 
These very faint candidates are indeed of relevant importance 
because their optical colours resemble those of ISO-CHA~II-13 
\citep{Alc06}. Note that also ISO-Cha~II-13 was not detected on 
our H$\alpha$ images. The H$\alpha$ information drawn from our data 
for the selected candidates is reported in Tab.~\ref{tab:par}.

%---------------------------------- Ha diag.-------------------
%-------------------------------------------------------------- 

\subsection{Towards the fraction of sub-stellar objects in ChaII}

An interesting and important quantity in this investigation
is the number of sub-stellar objects relative to the PMS stars 
in Cha~II. 
In order to determine this quantity, we must single out the sub-stellar 
objects. According to \citet{Cha00}, the sub-stellar limit for 3-4\,Myr 
objects, the approximate age of Cha~II \citep{Hug92, Alc00, Cie05}, 
falls at a temperature of about 2900\,K. Using the tools described 
in Sec.~\ref{analysis} and exploiting the potential of our photometric 
data, we attempted a first estimate of the temperature of the candidate 
members of Cha~II and hence, a guess on the fraction of sub-stellar 
candidates relative to the PMS stars. We derived first the dereddened 
SEDs of the candidates and then provided 
a first estimate of their temperature. 

%---------------------------------- SEDs-------------------
\begin{figure*} 
\centering
\includegraphics[width=14cm,height=21cm]{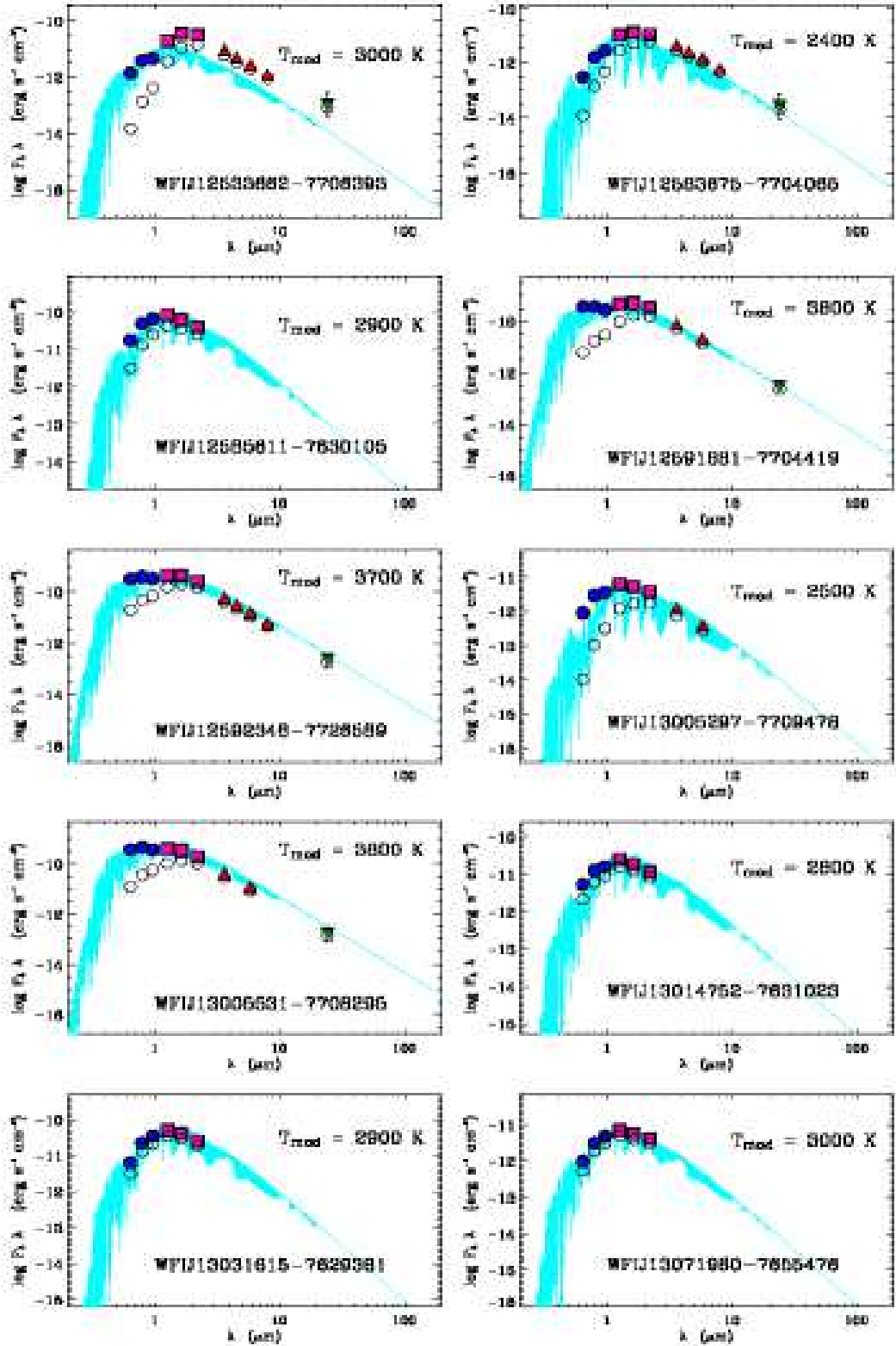}
\caption{Spectral energy distributions of the 10 new 
candidates selected in this work. Dereddened optical data are represented 
with filled circles; the IR data are from 2MASS (filled squares), 
IRAC@Spitzer and MIPS@Spitzer (filled triangles and upside down triangles 
respectively). The open circles represent the observed fluxes. 
The best fitting StarDusty spectra by \citet{All00} with the 
same temperature as the objects are over-plotted (see Sec.~\ref{Tlum}). 
WFI\,J12585611-7630105, WFI\,J13014752-7631023 and 
WFI\,J13031616-7629381 fall outside the areas mapped with 
MIPS and IRAC; although WFI\,J13071960-7655476 and 
WFI\,J13005297-7709478 fall in the area surveyed with MIPS, they were not detected.}
\label{sed_cand2}
\end{figure*}
%-------------------------------------------------------------- 

\subsubsection{Dereddened SEDs}
\label{seds}

The SEDs were derived by merging the $RIz$ WFI-Cousins photometry with 
the 2MASS $JHK$ magnitudes.  
The standard fluxes at each of the optical and near-IR pass-bands have 
been derived from the observed magnitudes using the expression  
$F_{\lambda} = F^0 _{\lambda} \cdot 10^{-0.4 \cdot mag^{corr} _{\lambda}}$,
where $mag^{corr} _{\lambda}$ is the observed magnitude corrected for 
interstellar reddening and $F^0 _{\lambda}$ is the Earth flux of an A0-type 
star of magnitude $V=0$ (Tab.~\ref{tab:flux0}). 
In order to derive the visual extinction, $A_V$, a SED minimization
procedure was used as follows. We assume that the observed magnitude at 
each wavelength ($m^{\lambda} _{obs}$) can be derived by the following 
equation:

\begin{equation}
m^{\lambda}_{obs} - m^{\lambda}_{ref} = A_V \cdot \frac{dA_\lambda}{dA_V} + S
\label{eq_fit}
\end{equation}

\noindent
where $m^{\lambda}_{ref}$ is a distance-scaled reference magnitude at 
a given wavelength ($\lambda$) and temperature (T$_{\rm eff}$). 
In Eq.~\ref{eq_fit} $A_V$ is the visual extinction, $A_\lambda$ the 
wavelength-dependent extinction and $S$ a scaling factor depending 
only on the stellar radius.
Using a grid of reference magnitudes and temperatures, we perform a 
linear fit to the previous expression were $\frac{dA_\lambda}{dA_V}$ 
and $m^{\lambda}_{obs} - m^{\lambda}_{ref}$ are the independent and 
dependent variables, respectively, and $A_V$ and $S$ the two free parameters 
of the fit. The $\frac{dA_\lambda}{dA_V}$ term was derived using the 
extinction law by \citet{Car89}, assuming the standard value $R_V$=3.1.
The grid of reference SEDs was constructed by combining the tabulations of colours 
as function of spectral types by \citet{Ken95}, \citet{Luh03} and \citet{Duc01}. 
The $A_V$ value of the reference SED minimizing the $\chi^2$ of the fit 
corresponds to the best approximation of the actual visual extinction. 
As a check, we have compared the results from our fitting procedure 
with the extinction derived from the spectral types and the observed 
colours of previously known PMS stars in Cha~II. 
The RMS difference in $A_V$ is less than 1.5\,mag, with only one object 
having a residual of about 4~mag.

\begin{table}
\caption[ ]{\label{tab:flux0} Absolute flux calibration constants and 
effective wavelengths for the optical and near-IR pass-bands.}
\begin{flushleft}
\small
\begin{tabular}{cccccc}  
\hline
Filter & $F^0 _{\lambda}$ & $\lambda_{\rm eff}$ & Ref. \\
       & (erg $\cdot$ s$^{-1}$ $\cdot$ cm$^{-2}$ $\cdot$ \AA$^{-1}$) & ($\mu$m) & \\
\noalign{\medskip}
\hline
\noalign{\medskip}
$R_C$ &	2.25 $\cdot 10^{-9}$   & 0.64  & \citet{Cou76} \\ 
$I_C$ & 1.19 $\cdot 10^{-9}$   & 0.79  & \citet{Cou76} \\ 
$z$   & 8.40 $\cdot 10^{-10}$  & 0.96  & This work     \\	    
$J$   & 3.13 $\cdot 10^{-10}$  & 1.25  & \citet{Cut03} \\	    
$H$   & 1.13 $\cdot 10^{-10}$  & 1.62  & \citet{Cut03} \\ 
$K$   & 4.28 $\cdot 10^{-11}$  & 2.20  & \citet{Cut03} \\ 
\noalign{\medskip}
\hline
\end{tabular}
\end{flushleft}
\end{table}

In Fig.~\ref{sed_cand2} the SEDs of the 10 new candidates selected 
in this work are shown. For six of these, we found a match in the
c2d Spitzer catalogues and their IRAC and/or MIPS fluxes are
also included. 
As appears from their SEDs, no significant IR excess is 
detected in these six objects. If they will be spectroscopically
confirmed as Cha~II members, they will most likely correspond to 
weak-line T Tauri stars (WTTS) or to BDs with thin disks. 
For the four remaining objects lacking Spitzer data we cannot
assess the presence of strong IR excess. However, for three of the 
latter we estimate an H$\alpha$ equivalenth width greater than about 20\AA.

It is interesting to note that the number of WTTS relative 
to CTTS in Cha~II is very small. Even assuming that all the 
candidates in our sample (see Tab.~\ref{tab:par}) would result 
in WTTS, which is not likely because several of them show 
evidence for strong H$\alpha$ emission, the CTTS 
would outnumber the WTTS by a factor $\sim$2. This would be in 
contrast with what it is found in other clouds, like Cha~I 
and Taurus, where WTTS dominate the PMS population \citep{Fei99}. 
Our selection based on the ($J-H$) vs. ($H-K$) diagram, picks up 
objects later than about K5. This might lead to the conclusion 
that some earlier type WTTS could have escaped our selection. 
However, such objects would have been detected in the previous
ROSAT X-ray surveys by \citet{Alc95, Alc00}. Only two WTTS 
in the cloud, namely RXJ1301.0-7654a and RXJ1303.1-7706 were 
discovered in such surveys. 
Thus, the low fraction of WTTS in Cha\,II seems to be real. 
More details on this will be discussed in a future c2d paper 
(Alcal\'a et al., in preparation), with a more complete set of 
SEDs for the Cha~II members.

%----------------------------------Teff comparison-------------------
\begin{figure} 
\resizebox{\hsize}{!}{\includegraphics[width=3cm,height=2.8cm]{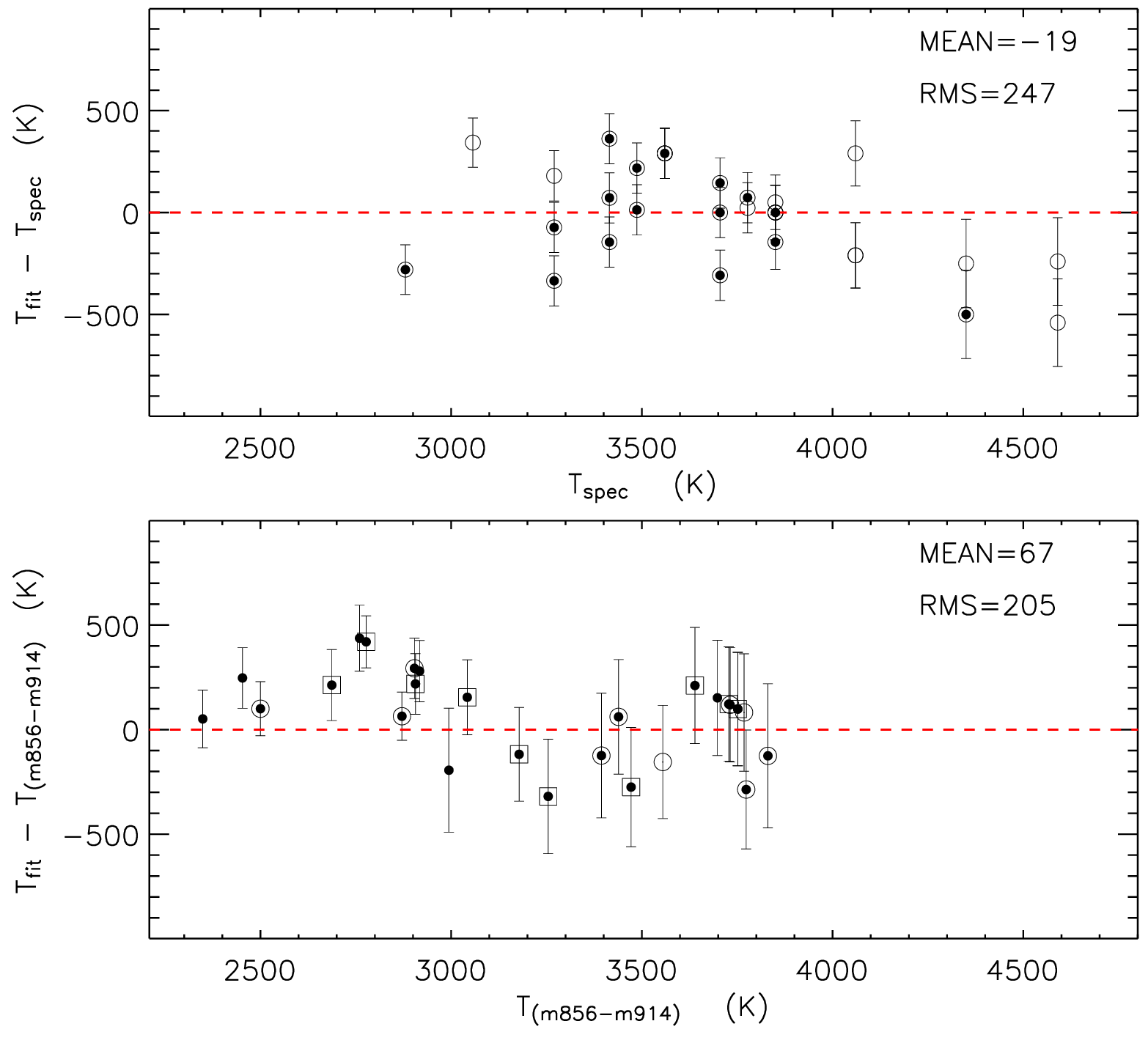}}
\caption{Upper panel: comparison between the effective temperature resulting 
from the SED fitting procedure described in Sec.~\ref{Tlum} and that derived 
from the spectral classification for the confirmed T~Tauri stars in Cha~II reported
by \citet{Hug92} and, for C~41, by \citet{Bar04} (open circles). 
The black dots mark the objects recovered by our selection criteria. 
The point at T$_{\rm spec}\approx$2900\,K
represents ISO-CHA\,II\,13 as reported by \citet{Alc06}. Lower Panel: comparison
between the effective temperature resulting from the SED fitting procedure
described in Sec.~\ref{Tlum} and that derived from the
T$_{\rm eff}$ vs. ($m_{856}-m_{914}$) calibration. The open circles and squares 
represent previously known PMS stars and candidates respectively. 
The black dots mark the objects recovered by our selection criteria}
\label{testT}
\end{figure}
%-------------------------------------------------------------- 

\subsubsection{Temperature}
\label{Tlum}

The temperature of the candidates was estimated using the 
T$_{\rm eff}$  vs. ($m_{856}-m_{914}$) calibration relation reported 
in Sec.~\ref{Teff}. This calibration is valid for cool objects with 
temperature in the range 2000--3800\,K, corresponding to a dereddened 
($m_{856}-m_{914}$) index in the range 1.16--0.16\,mag. Thus, we 
applied the calibration to objects with dereddened index 
($m_{856}-m_{914}$)$>$0.16\,mag. The $A_{\rm V}$ values determined as 
explained in Section~\ref{seds} and the extinction law by \citet{Car89}
were used to deredden the ($m_{856}-m_{914}$) index.
For objects with ($m_{856}-m_{914}$)$<$0.16\,mag, i.e. hotter than 
about 3800~K, a temperature estimate was done by fitting a grid of 
reference SEDs to the dereddened SEDs. The grid of reference SEDs was 
constructed by combining the tabulations of colours as a function of 
spectral type by \citet{Ken95}, \citet{Luh03} and \citet{Duc01}; the best approximation to the 
dereddened SED was then obtained by $\chi^2$-minimization of the 
flux differences between the dereddened SED and the reference SED. 
The minimization procedure was applied to the short-wavelength 
portion ($\lambda \leq \lambda_{J}$) of the SED, which is less 
affected by an eventual IR excess (Fig.~\ref{sed_cand2}).
The method results to be accurate within 250\,K relative to the 
spectroscopic temperature estimates for most of the previously 
known PMS stars.  
This is shown in the upper panel of Fig.~\ref{testT}, where the 
residuals between the temperature derived from the dereddened 
SED fitting, T$_{\rm fit}$, and that obtained from spectroscopy 
by \citet{Hug92}, T$_{\rm spec}$, are shown. For ISO-CHA\,II\,13 
we adopt the temperature value determined spectroscopically by 
\citet{Alc06}.

For the cool candidates, we also performed a consistency check 
between the temperature derived from SED fitting and that coming 
from the T$_{\rm eff}$ vs. ($m_{856}-m_{914}$) calibration.
We find that the two methods yield consistent results within 
$\sim$200\,K, as can be seen from the lower panel of Fig.~\ref{testT}. 
We can then use the T$_{\rm eff}$ vs. ($m_{856}-m_{914}$) 
calibration confidently for those objects with dereddened 
($m_{856}-m_{914}$)$>$0.16\,mag.

The resulting temperature for each of the candidates is reported
in Tab.~\ref{tab:par}. Using these estimates, we attempted to determine 
their radius and luminosity as follows:

\begin{enumerate}

\item 

First the reference SED of the same effective temperature 
as the object was selected and scaled to the Cha~II distance 
\citep[d$=178$ pc,][]{Whi97}:

 \begin{equation}
  F'_{ref}=F_{ref} \cdot \left ( \frac{R^\star}{d}  \right )^2
 \label{Fscale}
 \end{equation}

where R$^\star$ is the stellar radius; since R$^\star$ is an 
unknown quantity, we let it vary between 0.1, i.e. the value 
expected for a BD at the Deuterium burning limit \citep{Bar03}, 
and 10 R$_{\odot}$; for the objects whose effective temperature 
could not be estimated from the ($m_{856}-m_{914}$) index, 
both T$_{\rm eff}$ and R$^\star$ were set as free parameters 
in the assumption that the short-wavelength side of the dereddened 
SED ($\lambda \leq \lambda_{J}$) is mainly influenced by T$_{\rm eff}$;

\item 
the best estimate of R$^\star$ was then obtained by minimizing 
the flux differences between the observed dereddened SED and 
the reference one. The minimization procedure was performed, 
as above, to the short-wavelength portion 
($\lambda \leq \lambda_{J}$) of the SED;

\item 
 Finally, we calculated the bolometric 
 luminosities by applying  bolometric corrections to both 
 the dereddened $I$-band and $J$-band data. Bolometric corrections 
 were taken from \citet{Key95} for objects earlier than M6 
 (i.e. T$_{\rm eff} \gtrsim$3000~K), while for cooler objects we 
 followed the prescriptions by \citet{Luh99b} and combined the 
 compilations of bolometric corrections as a function of 
 temperature by \citet{Bes91}, \citet{Mon92}, \citet{Tin93} and \citet{Leg96}.

\end{enumerate}

Having an estimate of the temperature and luminosity of the 
candidates, we determined their masses and ages by comparison  
with the set of evolutionary tracks by \citet{Bar98} and \citet{Cha00}
on the HR diagram.\\

%--------------------- Table param --------------------------------

\begin{table*}
\caption[ ]{\label{tab:par} The 20 candidates selected in this work. 
The visual extinction (A$_{\rm V}$) and temperature (T$_{\rm eff}$) of the candidates 
were estimated as explained in Sec.~\ref{seds} and Sec.~\ref{Tlum}, respectively. Wherever possible, 
the extinction and temperature values were compared with other 
estimates from literature.}
\begin{center}
\footnotesize
\begin{tabular}{lccccccl}
\hline
 Designation & EW$_{H\alpha}$ & A$_{\rm V}$ & T$_{\rm eff}$ & A$^{lit} _V$   & T$^{lit} _{\rm eff}$  & Ref.  & Classification \\
             & ($\AA$)        & (mag)       &  (K)          & (mag)          &     (K)               &     & in this work \\
\noalign{\medskip}
\hline
\noalign{\medskip}
 WFI\,J12533662-7706393  &  Not Det.	       &  5.58  & $\dagger$3000$\pm$250	 &	 &	&    &  PMS star cand.  \\
	    C~17	 &  No Em. 	       &  5.95  & $\star$3600$\pm$200	 & 5.1   &	& b  &  PMS star cand.  \\
	    C~33	 &  No Em. 	       &  4.48  & $\dagger$3700$\pm$250	 & 2.4   &	& b  &  PMS star cand.  \\  
 IRAS~12535-7623	 &  No Em.  	       &  3.38  & $\star$3900$\pm$200	 & 5	 &	& a  &  PMS star cand.  \\     
 WFI\,J12583675-7704065  &  Not Det.	       &  3.84  & $\dagger$2400$\pm$200	 &	 &	&    &  BD cand.	\\    
 WFI\,J12585611-7630105  &  30 		       &  1.91  & $\dagger$2900$\pm$200	 &	 &	&    &  BD cand.        \\  
    ISO-CHA\,II\,29	 &  No Em.  	       &  5.94  & $\star$3900$\pm$200	 & 11.5  &	& b  &  PMS star cand.  \\  
 WFI\,J12591881-7704419  &  No Em.  	       &  4.98  & $\star$3800$\pm$200	 &	 &	&    &  PMS star cand.  \\  
 WFI\,J12592348-7726589  &  No Em.  	       &  3.33  & $\dagger$3700$\pm$250	 &	 &	&    &  PMS star cand.  \\  
 WFI\,J13005297-7709478  &  Not Det. 	       &  5.50  & $\dagger$2500$\pm$200	 &	 &	&    &  BD cand.	\\
 WFI\,J13005531-7708295  &  No Em.   	       &  4.13  & $\star$3800$\pm$200	 &	 &	&    &  PMS star cand.  \\     
  IRAS~F12571-7657	 &  No Em.   	       &  3.95  & $\star\star$$>$3000	 & 8,13  &	& b,c&  PMS star cand.  \\    
 WFI\,J13014752-7631023  &  No Em.   	       &  0.88  & $\dagger$2800$\pm$200	 &	 &	&    &  BD cand.	\\  
	    C~50	 &    20  	       &  3.46  & $\dagger$3000$\pm$200  & 2.0   &	& b  &  PMS star cand.  \\  
	    C~51	 &    20  	       &  1.16  & $\dagger$3300$\pm$250  & 2.3   &	& b  &  PMS star cand.  \\  
 WFI\,J13031615-7629381  &    15  	       &  0.61  & $\dagger$2900$\pm$200  &	 &	&    &  BD cand.        \\  
	   C~62 	 &    35  	       &  3.53  & $\dagger$3000$\pm$200  &4.1,4  & 3140 & b,c&  PMS star cand.  \\ 
 WFI\,J13071960-7655476  &    25  	       &  0.50  & $\star$3000$\pm$200	 &	 &	&    &  PMS star cand.  \\	 
	    C~66	 &    25  	       &  3.87  & $\dagger$2900$\pm$200  &3.7,0  & 2793 & b,c&  BD cand.	\\    
 2MASS13125238-7739182   &    10  	       &  1.36  & $\dagger$3400$\pm$250  &	 &	& d  &  PMS star cand.  \\  
\noalign{\medskip}
\hline
\end{tabular}
\end{center}
Notes to column 4:

$\dagger$ \footnotesize{Temperature derived from  the dereddened ($m_{856}-m_{914}$) colour index (Sec.~\ref{Teff}).}\\
$\star$ \footnotesize{Temperature derived from the SED fitting method described in Sec.~\ref{Tlum}.}\\
$\star\star$ \footnotesize{See Appendix~\ref{comm}.}\\

\footnotesize{References: a) \citet{Lar98}; b) \citet{Vuo01}; c) \citet{All06}; d) \citet{You05}.}
\end{table*}

\subsection{The fraction of sub-stellar objects}
\label{frac_sub}
 
Previous investigations \citep{Hug92,Alc00} showed some
evidences that the mass spectrum in Cha~II might be biased 
towards very-low mass objects. It is thus interesting to
investigate what our data suggest on this matter.

Assuming that all the 20 candidates are true cloud members, 
we end up with 6 candidates with T$_{\rm eff} \leq$2900\,K, 
i.e. below the substellar limit for 5\,Myr old objects. 
The remaining 14 objects result to be PMS star candidates (see Tab.~\ref{tab:par}). 
Taking into account the BD ISO-CHA\,II\,13 and the 33 confirmed 
PMS stars, the resulting fraction of sub-stellar objects ($\rm 0.02 M_{\odot} < M < 0.08 M_{\odot}$)
relative to the PMS stars ($\rm 0.08 M_{\odot} < M < 10 M_{\odot}$), $R_{ss}=\frac{N(0.02-0.08)}{N(0.08-10)}$, 
in Cha~II would be $\sim$15\% \citep[19\% if we consider the two planetary-mass objects reported by][]{Jay06}. 
This last value would imply a peculiar mass-spectrum for the population
of Cha~II relative to other T associations \citep[where $R_{ss}$=12-14\%,][]{Bri02,Lop04}, in particular in the 
sub-stellar domain. If we assume that only the candidates having
H$\alpha$ emission are true cloud members, the $R_{ss}$ value would drop 
to $\sim$10\% \citep[15\% if we consider the two planetary-mass objects reported by][]{Jay06}, 
i.e. more similar to that measured in other T associations and lower than 
that measured in OB associations \citep[$\sim$26\%,][]{Bou98,Hil00,Bar01,Bri02,Mue02}. 
If $R_{ss}$ in Cha~II is similar as in other T associations, we expect that at least 50\% 
of the candidates reported in Tab.~\ref{tab:par} will be genuine Cha~II members.

Assuming that the 7 IRAS sources with H$\alpha$ emission discussed in Appendix~\ref{comm} 
are PMS stars, the fraction of substellar
objects would drop to about 8\% or 13\% depending on whether 
the two planetary-mass objects are counted among the sub-stellar ones.

By comparing the position of the candidates in the HR diagram 
with the \citet{Bar98} and \citet{Cha00} evolutionary tracks, their 
age distribution would peak between 2 and 3 Myr, if they were 
genuine Cha~II members. This would be fairly consistent with the
average age of 3.6\,Myr determined by \citet{Cie05} for a 
sample of T~Tauri stars in Cha~II. Taking into account the 
candidates, the mean mass in Cha~II would be $\sim$0.5M$_{\odot}$, 
i.e. slightly lower than the mean mass of the confirmed 
members alone (0.6~M$_{\odot}$) and comparable 
to the mean mass for the Cha~I population  
($\sim$0.45 M$_{\odot}$\footnote{Luminosity and temperature determinations 
are reported by \citet{Luh04} for 144 members of Cha~I. We derived masses 
and ages for these objects using the evolutionary tracks by \citet{Bar98} 
and \citet{Cha00} in a homogeneous way.}). 

%\noindent
%Although our methods provide reasonable estimates of the physical 
%quantities that allow to draw some conclusions on the candidates, 
%more accurate determinations require also photometry in the IR 
%domain and spectroscopy in the optical and/or near IR. 

\section{Conclusions}
\label{sec:concl}

The optical survey presented here is one of the deepest and more 
extensive conducted so far in Cha~II and constitutes the optical 
ancillary data for the c2d Spitzer Legacy survey in this cloud. 

The photometric selection based on our optical imaging data, combined 
with data in the near-IR from the 2MASS, allowed us to recover basically 
all the previously known members of the cloud, including 10 candidates 
from previous IR surveys, with a high membership probability. 
Furthermore, we provided 10 new likely member candidates that were not 
detected by previous surveys, thereby increasing the total number of candidates to 20.
Should all these objects be spectroscopically confirmed 
as PMS stars and young BDs, the population of Cha~II will increase to 56 members. 
Up to 50\% of the sample may be contaminated by field dwarfs.
According to our characterisation criteria of the candidates, we estimate that 
at least some 50\% of them will result in true Cha~II members and, based on our 
temperature estimates, several of these objects are expected to be sub-stellar. 
Under these assumptions, we conclude that the fraction of substellar
objects relative to the PMS stars in Cha~II is on the order of $\sim$19\%, 
i.e. larger than that reported for other T associations like Taurus. 
In the most conservative hypothesis in which only the candidates showing 
H$\alpha$ emission would result in true member of the cloud, this fraction 
would drop to $\sim$15\%, i.e. comparable with that measured in other 
T associations.

From the completeness of our survey, both in space and flux, we 
conclude that the optical population of Cha~II members discovered 
so far is nearly complete. We do not expect to find many more PMS 
stars based on optical observations. 
More clues to these issues will be addressed in a forthcoming paper 
by combining the optical data presented here with those  %coming
from the MIPS@Spitzer and IRAC@Spitzer observations in Cha~II (Alcal\'a et al., in preparation).

\begin{acknowledgements}

This work was partially financed by the Istituto Nazionale 
di Astrofisica (INAF) and the Italian Ministero dell'Istruzione,
Universit\'a e Ricerca (MIUR). We thank the anonymous referee
for his/her constructive comments and suggestions. We also thank 
M. Radovich for many explanations on the use of ASTROMETRIX. 
We thank N. Evans, PI of the c2d Spitzer Legacy Program, and 
L. Cieza for many useful comments and suggestions on an earlier 
version of the paper. 
We also thank F. Com\'eron and E. Marilli for discussions. 
L. Spezzi acknowledges financial support from COFIN-MIUR-2004 
(The X-Shooter spectrograph for the VLT) and 
PRIN-INAF-2005 (Stellar clusters: a benchmark for star formation 
and stellar evolution) and a grant by the Europen Southern 
Observatory (ESO) for a two-months stay at Garching.  
Support for this work, part of the Spitzer Legacy Science 
Program, was provided by NASA through contract 1224608 issued 
by the Jet Propulsion Laboratory, California Institute of 
Technology, under NASA contract 1407. We also thank the Lorentz 
Center in Leiden for hosting several meetings that contributed 
to this paper. 
This research made use of the SIMBAD database,
operated at CDS (Strasbourg, France) and the data products from the Two-Micron 
All-Sky Survey, which is a joint project of the University 
of Massachusetts and the Infrared Processing and Analysis 
Center at California Institute of Technology, funded by NASA 
and the National Science Foundation.

\end{acknowledgements}

\bibliographystyle{aa} % style aa.bst

\begin{thebibliography}{}

\bibitem[\protect\citeauthoryear{Alcal\'{a} et al.}{1995}]{Alc95} Alcal\'{a} J.M., Krautter J., Schmitt J. et al. 1995, A\&A Supp. Ser. 114, 109
\bibitem[\protect\citeauthoryear{Alcal\'{a} et al.}{2000}]{Alc00} Alcal\'{a} J.M., Covino E., Sterzik M.F. et al. 2000, A\&A 355, 629
\bibitem[\protect\citeauthoryear{Alcal\'{a} et al.}{2002}]{Alc02} Alcal\'{a} J.M., Radovich M., Silvotti R. et al. 2002, proceedings of the SPIE 4836, 406
\bibitem[\protect\citeauthoryear{Alcal\'{a} et al.}{2004}]{Alc04} Alcal\'{a} J.M., Wachter S., Covino E. et al. 2004, A\&A 416, 677
\bibitem[\protect\citeauthoryear{Alcal\'{a} et al.}{2006}]{Alc06} Alcal\'{a} J.M., Spezzi, L., Frasca, A., Covino, E., Porras, A., Mer\'{i}n, B., Persi, P. 2006, A\&A, 453, 1
\bibitem[\protect\citeauthoryear{Allard}{1990}]{All90} Allard F. 1990, PhD Thesis, Ruprecht Karls Univ. Heidelberg
\bibitem[\protect\citeauthoryear{Allard et al.}{2000}]{All00} Allard F., Hauschildt P.H. \& Schwenke D. 2000, ApJ 540, 1005
\bibitem[\protect\citeauthoryear{Allard et al.}{2001}]{All01} Allard F., Hauschildt P.H., Alexander D.R. et al. 2001, ApJ 556, 357
\bibitem[\protect\citeauthoryear{Allers et al.}{2006}]{All06} Allers K.N., Kessler-Silacci J.E., Cieza L.A. \& Jaffe D.T. 2006, ApJ 644, 364 
\bibitem[\protect\citeauthoryear{Baraffe et al.}{1998}]{Bar98} Baraffe I., Chabrier G., Allard F. \& Hauschildt P.H. 1998, A\&A 337, 403 
\bibitem[\protect\citeauthoryear{Baraffe et al.}{2002}]{Bar02} Baraffe I., Chabrier G., Allard F. \& Hauschildt P.H. 2002, A\&A 382, 563
\bibitem[\protect\citeauthoryear{Baraffe et al.}{2003}]{Bar03} Baraffe I., Chabrier G., Barman T.S. et al. 2003, A\&A 402, 701
\bibitem[\protect\citeauthoryear{Barrado y Navascu\`{e}s et al.}{2001}]{Bar01} Barrado Y Navascu\`{e}s D., Stauffer J.R., Briceno C. et al. 2001, ApJS 134, 103
\bibitem[\protect\citeauthoryear{Barrado y Navascu\`{e}s \& Jayawardhana}{2004}]{Bar04} Barrado Y Navascu\`{e}s D. \& Jayawardhana R. 2004, ApJ 615, 840
\bibitem[\protect\citeauthoryear{Bessel}{1990}]{Bes90} Bessel M.S., PASP 102, 1181
\bibitem[\protect\citeauthoryear{Bessel}{1991}]{Bes91} Bessell M.S. 1991, AJ 101, 662
\bibitem[\protect\citeauthoryear{Bohlin \& Gilliland}{2004}]{Boh04} Bohlin R.C. \& Gilliland R.L 2004, ApJ 128, 3053
\bibitem[\protect\citeauthoryear{Boulanger et al.}{1998}]{Bou98} Boulanger F., Bronfman L., Dame T.M. \& Thaddeus P. 1998, A\&A 332, 273
\bibitem[\protect\citeauthoryear{Brice\~no et al.}{2002}]{Bri02} Brice\~no, Luhman K.L., Hartmann L. et al. 2002, ApJ 580, 317
\bibitem[\protect\citeauthoryear{Burgasser}{2004}]{Bur04a} Burgasser A.J. 2004,ApJ Supp. Ser. 155, 207
\bibitem[\protect\citeauthoryear{Burgasser et al.}{2004}]{Bur04b} Burgasser A.J., Kirkpatrick J.D., McGovern M.R. et al. 2004, ApJ 604, 827
\bibitem[\protect\citeauthoryear{Burrows \& Liebert}{1993}]{Bur93} Burrows A. \& Liebert J. 1993, Rev. of Modern Physics Vol. 65, No. 2, p. 301
\bibitem[\protect\citeauthoryear{Cambr\'esy}{1999}]{Cam99} Cambr\'esy L. 1999, A\&A 345, 965
\bibitem[\protect\citeauthoryear{Cardelli et al.}{1989}]{Car89} Cardelli J.A., Clayton G.C. \& Mathis J.S. 1989, ApJ 345, 245
\bibitem[\protect\citeauthoryear{Chabrier et al.}{2000}]{Cha00} Chabrier G., Baraffe I., Allard F. \& Hauschildt P. 2000, ApJ 542, 464
\bibitem[\protect\citeauthoryear{Chen et al.}{1995}]{Che95} Chen H., Myers P.C., Ladd E.F. \& Wood D.O.S. 1995, ApJ 445, 377
\bibitem[\protect\citeauthoryear{Chen et al.}{1997}]{Che97} Chen H., Grenfell T.G., Myers P.C. \& Hughes J.D. 1997, ApJ 478, 295
\bibitem[\protect\citeauthoryear{Chen et al.}{2001}]{Che01} Chen B.C., M\'{e}ndez, Ren\'{e} A., Tsay W.S. \& Lu P.K. 2001, ApJ 121, 309
\bibitem[\protect\citeauthoryear{Cieza et al.}{2005}]{Cie05} Cieza, L. A., Kessler-Silacci, J. E., Jaffe, D.T., et al. 2005, ApJ, 635, 422  
\bibitem[\protect\citeauthoryear{Cousins}{1976}]{Cou76} Cousins A.W.J. 1976, Mem. R. Astr. Soc. 81, 25
\bibitem[\protect\citeauthoryear{Covino et al.}{1997}]{Cov97} Covino E., Alcal\'{a} J. M., Allain S. et al. 1997, A\&A 328, 187
\bibitem[\protect\citeauthoryear{Cutri et al.}{2003}]{Cut03} Cutri R.M., Skrutskie M.F., Van Dyk S. et al. 2003, \emph{Explanatory Supplement to the 2MASS All Sky Data Release}
\bibitem[\protect\citeauthoryear{Dahn et al.}{2002}]{Dah02} Dahn C.C., Harris  H.C., Vrba F.J. et al. 2002, AJ 124, 1170
\bibitem[\protect\citeauthoryear{D'Antona \& Mazzitelli}{1998}]{DAn98} D'Antona F. \& Mazzitelli I. 1997, Mem. S.A.It. 68, 807
\bibitem[\protect\citeauthoryear{Ducati et al.}{2001}]{Duc01} Ducati J.R., Bevilacqua C.M., Rembold S.B. \& Ribeiro D. 2001, ApJ 558, 309
\bibitem[\protect\citeauthoryear{Dullemond et al.}{2001}]{Dul01} Dullemond C.P., Dominik C. \& Natta A. 2001, ApJ 560, 957
\bibitem[\protect\citeauthoryear{Evans et al.}{2003}]{Eva03} Evans N.J. II, Allen L.E., Blake G.A. et al. 2003, PASP 115, 965
\bibitem[\protect\citeauthoryear{Feigelson et al.}{1993}]{Fei93} Feigelson E.D., Casanova S., Montmerle T. \& Guibert J. 1993, ApJ 416, 623
\bibitem[\protect\citeauthoryear{Feigelson \& Montmerle}{1999}]{Fei99} Feigelson E.D. \& Montmerle T. 1999, Ann. Rev. A\&A 37, 363
\bibitem[\protect\citeauthoryear{Fern\`andez et al.}{1995}]{Fer95} Fern\`andez M., Ortiz E., Eiroa C. \& Miranda L.F. 1995, A\&A Supp. Ser. 114, 439
\bibitem[\protect\citeauthoryear{Garay et al.}{2002}]{Gar02} Garay G., Mardones D., Rodr\`iguez L.F. et al. 2002, ApJ 567, 980
\bibitem[\protect\citeauthoryear{Giannini et al.}{2006}]{Gia06} Giannini T., Mccoey C., Nisini B. et al. 2006, A\&A, in press
\bibitem[\protect\citeauthoryear{Graham \& Hartigan}{1988}]{Gra88} Graham J.A. \& Hartigan P. 1988, ApJ 95, 1197
\bibitem[\protect\citeauthoryear{Gray}{1992}]{Gra92} Gray D.F. 1992, \emph{The observations and analysis of stellar photospheres}, Cambridge Univ. Press
\bibitem[\protect\citeauthoryear{Guti\`{e}rrez Moreno et al.}{1988}]{Gut88} Guti\`{e}rrez Moreno A., Moreno H., Cortes G. \& Wenderoth E. 1988, PASP 100, 973
\bibitem[\protect\citeauthoryear{Hamuy et al.}{1992}]{Ham92} Hamuy M., Walker A.R., Suntzeff N.B. et al. 1992, PASP 104, 533
\bibitem[\protect\citeauthoryear{Hartigan}{1993}]{Har93} Hartigan P. 1993 ApJ 105, 1511
\bibitem[\protect\citeauthoryear{Hauschildt et al.}{1999}]{Hau99} Hauschildt P.H., Allard F. \& Baron E. 1999, ApJ 512, 377
\bibitem[\protect\citeauthoryear{Henning et al.}{1993}]{Hen93} Henning T., Pfau W., Zinnecker H. \& Prusti T. 1993, A\&A 276, 129
\bibitem[\protect\citeauthoryear{Hillenbrand \& Carpenter}{2000}]{Hil00} Hillenbrand L.A. \& Carpenter J. 2000, ApJ 540, 236
\bibitem[\protect\citeauthoryear{Hughes \& Hartigan}{1992}]{Hug92} Hughes J.H. \& Hartigan P. 1992, ApJ 104, 680
\bibitem[\protect\citeauthoryear{Jacoby et al.}{1987}]{Jac87} Jacoby G.H., Africano J.L. \& Quigley R.J. 1987, PASP 99, 672
\bibitem[\protect\citeauthoryear{Jayawardhana \& Ivanov}{2006}]{Jay06} Jayawardhana R. \& Ivanov V.D. 2006, in press
\bibitem[\protect\citeauthoryear{Jones et al.}{1981}]{Jon81} Jones T., Ashley M., Hyland A. \& Ruelas-Mayoroga A. 1981, MNRAS 197, 413\bibitem[\protect\citeauthoryear{Kenyon \& Hartmann}{1995}]{Key95} Kenyon S.J. \& Hartmann L. 1995, ApJ Supp. Ser. 101, 117
\bibitem[\protect\citeauthoryear{Kenyon \& Hartmann}{1995}]{Ken95} Kenyon S.J. \& Hartmann L. 1995, ApJ Supp. Ser. 101, 117
\bibitem[\protect\citeauthoryear{Kroupa}{2001}]{Kro01} Kroupa P. 2001, Mon. Not. R. Astron. Soc. 322, 231
\bibitem[\protect\citeauthoryear{Kroupa}{2002}]{Kro02} Kroupa P. 2002, Science 295, 82
\bibitem[\protect\citeauthoryear{Kurucz}{1979}]{Kur79} Kurucz R.L. 1979, ApJ Suppl. Ser. 40, 191
\bibitem[\protect\citeauthoryear{Landolt}{1992}]{Lan92} Landolt A.U. 1992, ApJ, 104, 340
\bibitem[\protect\citeauthoryear{Larson et al.}{1998}]{Lar98} Larson K.A., Whittet D.C.B., Prusti T. \& Chiar J.E. 1998, A\&A 337, 465
\bibitem[\protect\citeauthoryear{Lee et al.}{2005}]{Lee05} Lee H.T., Chen W.P., Zhi W.Z. \& Jing Y.H. 2005, ApJ 624, 808
\bibitem[\protect\citeauthoryear{Leggett et al.}{1996}]{Leg96} Leggett S.K., Allard F.,Berriman G. et al. 1996, ApJ Supp. Ser. 104, 117
\bibitem[\protect\citeauthoryear{L\'opez Mart\'i et al.}{2004}]{Lop04} L\'opez Mart\'i B., Eisl\"{o}ffel J., Scholz A. \& Mundt R. 2004, A\&A 416, 555
\bibitem[\protect\citeauthoryear{L\'opez Mart\'i et al.}{2005}]{Lop05} L\'opez Mart\'i B., Eisl\"{o}ffel J. \& Mundt R. 2005, A\&A 444, 175
\bibitem[\protect\citeauthoryear{Luhman \& Rieke}{1999}]{Luh99a} Luhman K.L. \& Rieke G.H. 1999, ApJ 525, 440
\bibitem[\protect\citeauthoryear{Luhman}{1999}]{Luh99b} Luhman K.L. 1999, ApJ 525, 466
\bibitem[\protect\citeauthoryear{Luhman et al.}{2000}]{Luh00} Luhman K.L., Rieke G.H., Young E.T. et al. 2000, ApJ 540, 1016
\bibitem[\protect\citeauthoryear{Luhman et al.}{2003}]{Luh03} Luhman K.L., Stauffer J.R., Muench A.A. et al. 2003, ApJ 593,1093
\bibitem[\protect\citeauthoryear{Luhman}{2004}]{Luh04} Luhman K.L. 2004, ApJ 602, 816
\bibitem[\protect\citeauthoryear{Luhman et al.}{2005}]{Luh05} Luhman K.L., D'Alessio P., Calvet N. et al. 2005, ApJ, 620, 51
\bibitem[\protect\citeauthoryear{Meyer et al.}{1997}]{Mey97} Meyer M.R., Calvet N. \& Hillenbrand L.A. 1997, ApJ 114, 288
\bibitem[\protect\citeauthoryear{Minezaki \& Kobayashi}{1998}]{Min98} Minezaki T. \& Kobayashi Y. 1998, ApJ 494, 111
\bibitem[\protect\citeauthoryear{Monet et al.}{1992}]{Mon92} Monet D.G., Dahn C.C., Vrba F.J. et al. 1992, AJ 103, 638
\bibitem[\protect\citeauthoryear{Monet et al.}{1998}]{Mon98} Monet D., Bird A., Canzian B. et al. 1998, \emph{The USNO-A2.0 catalogue}, U.S. Naval Observatory
\bibitem[\protect\citeauthoryear{Muench et al.}{2002}]{Mue02} Muench A.A., Lada E.A., Lada C.J. \& Alves J. 2002, ApJ 573, 366
\bibitem[\protect\citeauthoryear{Muench et al.}{2003}]{Mue03} Muench A.A., Lada E.A., Lada C.J.  et al. 2003, ApJ 125, 2029
\bibitem[\protect\citeauthoryear{O'Neal et al.}{1998}]{ONe98} O'Neal D., Neff J.E. \& Saar S.H. 1998, ApJ 507, 919
\bibitem[\protect\citeauthoryear{Palla \& Stahler}{1999}]{Pal99} Palla F. \& Stahler S.W. 1999, ApJ 525, 772
\bibitem[\protect\citeauthoryear{Persi et al.}{2003}]{Per03} Persi P., Marenzi A. R., G\'{o}mez M. \& Olofsson G. 2003, A\&A 399, 995
\bibitem[\protect\citeauthoryear{Porras et al.}{2006}]{Por06} Porras A., Jorgensen, J.K, Allen, L. et al., 2006 submitted
\bibitem[\protect\citeauthoryear{Preibisch et al.}{2003}]{Pre03} Preibisch T., Stanke T. \& Zinnecker H. 2003, A\&A  409, 147
\bibitem[\protect\citeauthoryear{Preibisch et al.}{2005}]{Pre05} Preibisch T., McCaughrean M.J., Grosso N. et al. 2005, ApJ Supp. Ser. 160, 582
\bibitem[\protect\citeauthoryear{Prescott et al.}{2006}]{Pre06} Prescott, M.K.M., Impey, C.D., Cool, R.J., Scoville, N. 2006, ApJ 644, 100
\bibitem[\protect\citeauthoryear{Prusti et al.}{1992}]{Pru92} Prusti T., Whittet D.C.B., Assendorp R. \& Wesselius P.R. 1992, A\&A 260, 151
\bibitem[\protect\citeauthoryear{Reipurth \& Clarck}{2001}]{Rei01} Reipurth B. \& Clarck C. 2001, ApJ 122, 432
\bibitem[\protect\citeauthoryear{Schwartz}{1977}]{Sch77} Schwartz R.D. 1977, ApJS 35,161
\bibitem[\protect\citeauthoryear{Schwartz}{1991}]{Sch91} Schwartz R.D. 1991, ESO rep. No. 11, p. 93
\bibitem[\protect\citeauthoryear{Stetson}{1987}]{Ste87} Stetson P.B. 1987, PASP 99, 191
\bibitem[\protect\citeauthoryear{Stetson}{2000}]{Ste00} Stetson P.B. 2000, PASP 112, 925
\bibitem[\protect\citeauthoryear{Tinney et al.}{1993}]{Tin93} Tinney C.G., Mould J.R. \& Reid I.N. 1993, AJ 105, 1045
\bibitem[\protect\citeauthoryear{Vuong et al.}{2001}]{Vuo01} Vuong M.H., Cambr\'{e}sy L. \& Epchtein N. 2001, A\&A 379, 208
\bibitem[\protect\citeauthoryear{White \& Basri}{2003}]{Whi03} White R.J.\& Basri G. 2003, ApJ 582, 1109
\bibitem[\protect\citeauthoryear{Whittet et al.}{1991}]{Whi91} Whittet D.C.B., Laureijs R.J., Zhang C.Y. 1991, A\&A 251, 524
\bibitem[\protect\citeauthoryear{Whittet et al.}{1997}]{Whi97} Whittet D.C.B., Prusti T., Franco G.A.P. et al. 1997, A\&A 327, 1194
\bibitem[\protect\citeauthoryear{Young et al.}{2005}]{You05} Young K.E., Harvey P.M., Brooke T.Y. et al. 2005, ApJ 628, 283
\bibitem[\protect\citeauthoryear{Zerbi et al.}{2006}]{Zer06} Zerbi F.M., Pallavicini R., Conconi P. et al. 2006, Mem. S.A.It. Suppl., Vol. 9, 419


\end{thebibliography}

\Online

%--------------------- Table photometry --------------------------------

\newpage

\topmargin 2cm
\pagestyle{empty}
	
\begin{landscape}
\scriptsize
\begin{longtable}{lccccccccccl}
\caption[ ]{\label{tab:phot} Optical photometry of objects in the Cha~II dark cloud. The positions 
for most objects are from the $R$-band images. For saturated objects in the $R$-band the positions are 
from the $H\alpha_{12}$ images; the coordinates of Sz~62, not observed in H$\alpha$, are from the 
$z$-band image. 
Magnitudes marked with an asterisk are taken from the the literature (see the relative references) 
and correspond to objects saturated in our images. Symbols and labels are as explained in the footnote. 
Comments about some of these objects are given in Appendix~\ref{comm}.}\\
\hline
Designation & RAJ2000        & DECJ2000       & $R_{\rm c}$ & $I_{\rm c}$ & $z$ & $H\alpha_{7}$ & $H\alpha_{12}$ & $m_{856}$ & $m_{914}$ & Main Ref. & Note \\ 
            & (hh:mm:ss)     & (dd:mm:ss)     &	      &	            &     & 	          & 	           &	       &	   &           &            \\ 
\hline
\endfirsthead
\caption{Continued.}\\
\hline
Designation & R.A.$_{J2000}$ & Dec.$_{J2000}$ & $R_{\rm c}$ & $I_{\rm c}$ & $z$ & $H\alpha_{7}$ & $H\alpha_{12}$ & $m_{856}$ & $m_{914}$ & Main Ref. & Note \\ 
	     & (hh:mm:ss)     & (dd:mm:ss)     &		&	      &     &		    &		     &  	 &	     &      &	   \\ 
\hline
\endhead
\hline
\endfoot
	IRAS~12416-7703   & 12:45:06.43 & $-$77:20:13.52 &    S 		&	 S	       &    S		& 11.86$\pm$0.09 & 12.10$\pm$0.13 &   S 	   &   S	    & i   & $\bullet$ 		      	 	\\
	IRAS~12448-7650   & 12:48:25.70 & $-$77:06:36.72 &    S 		&	 S	       &    S		& 13.15$\pm$0.09 & 13.55$\pm$0.13 &   S 	   &   S	    & i   & $\bullet$ 		      	 	\\
       IRAS~F12488-7658   & 12:52:30.49 & $-$77:15:12.92 & 14.84$\pm$0.01	& 12.21$\pm$0.02$^\ast$&    S		& 15.48$\pm$0.10 & 15.60$\pm$0.13 & 12.27$\pm$0.09 &   S	    & f,i & $\bullet$ 		      	 	\\
	IRAS~12496-7650   & 12:53:17.17 & $-$77:07:10.63 & 16.43$\pm$0.01	& 13.92$\pm$0.01       & 12.82$\pm$0.02 & 16.12$\pm$0.10 & 16.84$\pm$0.14 & 13.52$\pm$0.10 & 13.41$\pm$0.08 & a   & $\bigstar$ \leftthumbsup  	 	\\
    WFI\,J12533662-7706393& 12:53:36.62 & $-$77:06:39.31 & 22.22$\pm$0.14	& 19.35$\pm$0.04       & 18.00$\pm$0.04 &	  ND	 &	     ND   & 19.26$\pm$0.20 & 18.58$\pm$0.12 & m   & \leftthumbsup$^{NEW}$        	\\   
		   C~17   & 12:53:38.84 & $-$77:15:53.21 & 17.08$\pm$0.01	& 15.09$\pm$0.01       & 14.23$\pm$0.01 & 17.67$\pm$0.10 & 17.36$\pm$0.13 & 15.10$\pm$0.09 & 14.79$\pm$0.08 & f   & $\bullet$ \leftthumbsup      	\\  
	IRAS~12500-7658   & 12:53:42.79 & $-$77:15:11.59 & 21.54$\pm$0.08	& 19.40$\pm$0.01       & 18.44$\pm$0.06 & 20.81$\pm$0.38 & 21.24$\pm$0.46 & 19.43$\pm$0.23 & 19.06$\pm$0.15 & i   & $\bigstar$ \ding{74}      	 	\\
	      ChaII~304   & 12:55:16.00 & $-$76:46:21.83 & 21.85$\pm$0.09	& 19.28$\pm$0.02       & 18.27$\pm$0.05 &	ND	 &	  ND	  & 19.26$\pm$0.17 & 18.65$\pm$0.11 & j   & $\bullet$ \ding{68} \leftthumbsdown \\
	      ChaII~305   & 12:55:16.48 & $-$76:46:20.89 & 21.82$\pm$0.09	& 19.30$\pm$0.01       & 18.29$\pm$0.05 &	ND	 &	  ND	  & 19.20$\pm$0.17 & 18.68$\pm$0.12 & j   & $\bullet$ \ding{68} \leftthumbsdown \\
		   C~33   & 12:55:25.72 & $-$77:00:46.62 & 16.86$\pm$0.01	& 15.16$\pm$0.01       & 14.66$\pm$0.01 & 17.15$\pm$0.22 & 16.93$\pm$0.13 & 15.22$\pm$0.09 & 14.79$\pm$0.08 & f   & $\bullet$ \leftthumbsup  		\\
 2MASS12560549-7654106    & 12:56:05.43 & $-$76:54:10.69 & 15.95$\pm$0.01  	& 14.54$\pm$0.01       & 14.04$\pm$0.02 & 16.17$\pm$0.10 & 16.15$\pm$0.13 & 14.58$\pm$0.09 & 14.42$\pm$0.08 & i   & $\bullet$ \leftthumbsdown 		\\
		Sz~46NW   & 12:56:32.85 & $-$76:45:44.78 & 18.45$\pm$0.01	& 17.69$\pm$0.17       & 17.56$\pm$0.04 & 18.72$\pm$0.13 & 18.70$\pm$0.14 & 17.96$\pm$0.10 & 18.07$\pm$0.09 & a   & $\bullet$ \ding{68} \leftthumbsdown \\
		 Sz~46N   & 12:56:33.59 & $-$76:45:45.18 & 14.61$\pm$0.01	& 13.16$\pm$0.01       & 12.68$\pm$0.01 & 14.80$\pm$0.10 & 15.07$\pm$0.13 & 13.24$\pm$0.09 & 13.08$\pm$0.08 & a   & $\bigstar$ \ding{68} \leftthumbsup  \\
		 Sz~46S   & 12:56:33.64 & $-$76:45:49.54 & 16.81$\pm$0.01	& 16.05$\pm$0.02       & 15.93$\pm$0.02 & 17.07$\pm$0.10 & 17.03$\pm$0.13 & 16.32$\pm$0.09 & 16.38$\pm$0.08 & a   & $\bullet$ \ding{68} \leftthumbsdown \\
		  Sz~47   & 12:56:58.63 & $-$76:47:06.72 & 13.86$\pm$0.01$^\ast$& 13.38$\pm$0.01$^\ast$& 13.45$\pm$0.01 & 13.68$\pm$0.10 & 14.07$\pm$0.13 & 13.72$\pm$0.09 & 13.86$\pm$0.08 & a   & $\bigstar$ \ding{80}        	\\
	IRAS~12533-7632   & 12:57:00.49 & $-$76:48:35.10 & 21.19$\pm$0.12	& 19.76$\pm$0.01       &       ND	&	ND	 & 20.90$\pm$0.38 & 19.92$\pm$0.33 & 19.83$\pm$0.29 & e   & $\bullet$		        	\\
	IRAS~12535-7623   & 12:57:11.64 & $-$76:40:11.14 & 13.61$\pm$0.01	& 12.13$\pm$0.01       & 11.53$\pm$0.02 & 13.69$\pm$0.10 & 13.73$\pm$0.13 & 12.22$\pm$0.09 & 12.03$\pm$0.08 & e,i & $\bullet$ \leftthumbsup     	\\
	   ISO-CHA\,II\,13& 12:58:06.67 & $-$77:09:09.22 & 22.51$\pm$0.27	& 19.38$\pm$0.01       & 17.89$\pm$0.04 &	ND	 &	 ND	  & 19.44$\pm$0.19 & 18.33$\pm$0.11 & g,k & $\bigstar$ \leftthumbsup    	\\
    WFI\,J12583675-7704065& 12:58:36.75 & $-$77:04:06.53 & 22.45$\pm$0.25	& 19.32$\pm$0.01       & 17.85$\pm$0.04 &	ND	 &	  ND	  & 19.51$\pm$0.20 & 18.27$\pm$0.10 & m   & \leftthumbsup$^{NEW}$       	\\
    WFI\,J12585611-7630105& 12:58:56.11 & $-$76:30:10.48 & 16.39$\pm$0.01	& 14.38$\pm$0.01       & 13.54$\pm$0.02 & 16.08$\pm$0.09 & 16.60$\pm$0.13 & 14.34$\pm$0.09 & 13.81$\pm$0.08 & m   & \leftthumbsup$^{NEW}$       	\\
		   C~41   & 12:59:09.86 & $-$76:51:03.49 & 18.51$\pm$0.01	& 17.03$\pm$0.01       & 16.45$\pm$0.02 & 18.06$\pm$0.10 & 18.88$\pm$0.14 & 17.31$\pm$0.09 & 17.07$\pm$0.08 & f,h & $\bigstar$ \ding{80}        	\\
	   ISO-CHA\,II\,29& 12:59:10.19 & $-$77:12:13.72 & 16.19$\pm$0.02	& 14.23$\pm$0.01       & 13.36$\pm$0.01 & 16.41$\pm$0.19 & 16.32$\pm$0.13 & 14.37$\pm$0.09 & 13.91$\pm$0.08 & g,i & $\bullet$ \leftthumbsup     	\\
    WFI\,J12591881-7704419& 12:59:18.81 & $-$77:04:41.92 & 15.69$\pm$0.01       & 14.09$\pm$0.01       & 13.34$\pm$0.01 & 15.86$\pm$0.10 & 15.88$\pm$0.17 & 14.19$\pm$0.09 & 13.86$\pm$0.09 & m   & \leftthumbsup$^{NEW}$		\\
    WFI\,J12592348-7726589& 12:59:23.48 & $-$77:26:58.96 & 14.41$\pm$0.01       & 12.99$\pm$0.01       & 12.39$\pm$0.01 & 14.78$\pm$0.12 & 14.62$\pm$0.12 & 13.13$\pm$0.09 & 12.75$\pm$0.08 & m   & \leftthumbsup$^{NEW}$		\\
	IRAS~12556-7731   & 12:59:26.50 & $-$77:47:08.70 &     S		&   S		       &       S	& 12.34$\pm$0.09 & 12.66$\pm$0.13 &   S 	   &   S	    & i   & $\bullet$		        	\\
    WFI\,J13005297-7709478& 13:00:52.97 & $-$77:09:47.77 & 22.63$\pm$0.30	& 19.64$\pm$0.07       & 18.25$\pm$0.05 &	  ND	 &	     ND   & 19.85$\pm$0.25 & 18.66$\pm$0.12 & m   & \leftthumbsup$^{NEW}$       	\\ 
		Sz~48NE   & 13:00:53.15 & $-$77:09:09.18 & 16.08$\pm$0.01	& 14.32$\pm$0.04       & 13.55$\pm$0.01 & 16.11$\pm$0.09 & 16.24$\pm$0.13 & 14.44$\pm$0.09 & 14.02$\pm$0.08 & a   & $\bigstar$ \ding{68} \leftthumbsup  \\
		  Sz~49   & 13:00:53.26 & $-$76:54:14.98 & 14.86$\pm$0.01	& 13.53$\pm$0.03       & 13.01$\pm$0.01 & 13.90$\pm$0.09 & 14.93$\pm$0.13 & 13.62$\pm$0.09 & 13.35$\pm$0.08 & a   & $\bigstar$  \ding{80}     		\\
		Sz~48SW   & 13:00:53.56 & $-$77:09:08.28 & 16.26$\pm$0.01	& 14.50$\pm$0.02       & 13.72$\pm$0.01 & 16.13$\pm$0.09 & 16.57$\pm$0.13 & 14.51$\pm$0.09 & 14.19$\pm$0.08 & a   & $\bigstar$ \ding{68}\leftthumbsup   \\
		  Sz~50   & 13:00:55.28 & $-$77:10:22.01 & 14.42$\pm$0.01	& 12.50$\pm$0.01$^\ast$& 11.96$\pm$0.02 & 14.34$\pm$0.10 & 14.74$\pm$0.13 & 12.79$\pm$0.09 & 12.43$\pm$0.08 & a   & $\bigstar$ \leftthumbsup    	\\
    WFI\,J13005531-7708295& 13:00:55.31 & $-$77:08:29.54 & 14.93$\pm$0.01	& 13.29$\pm$0.01       & 12.64$\pm$0.01 & 15.26$\pm$0.14 & 15.24$\pm$0.13 & 13.39$\pm$0.09 & 13.15$\pm$0.09 & m   & \leftthumbsup$^{NEW}$       	\\
	RXJ1301.0-7654a   & 13:00:56.22 & $-$76:54:01.76 & 11.59$\pm$0.01$^\ast$& 10.54$\pm$0.01$^\ast$& 10.17$\pm$0.01 & 11.63$\pm$0.09 & 11.71$\pm$0.14 & 10.61$\pm$0.09 & 10.66$\pm$0.09 & d   & $\bigstar$\leftthumbsup     	\\
       IRAS~F12571-7657   & 13:00:59.21 & $-$77:14:02.80 & 18.62$\pm$0.01	& 16.16$\pm$0.01       & 15.00$\pm$0.01 & 19.26$\pm$0.14 & 19.23$\pm$0.16 & 16.48$\pm$0.23 & 15.79$\pm$0.14 & e,g & $\bullet$ \leftthumbsup     	\\
	   ISO-CHA\,II\,73& 13:01:46.03 & $-$77:16:02.89 & 16.31$\pm$0.02	& 15.12$\pm$0.01       & 14.68$\pm$0.01 & 16.71$\pm$0.22 & 16.55$\pm$0.13 & 15.38$\pm$0.09 & 15.13$\pm$0.08 & g   & $\bullet$ \leftthumbsdown   	\\
    WFI\,J13014752-7631023& 13:01:47.52 & $-$76:31:02.32 & 16.83$\pm$0.05	& 15.23$\pm$0.01       & 14.65$\pm$0.06 & 16.84$\pm$0.20 & 17.05$\pm$0.21 & 15.77$\pm$0.17 & 15.18$\pm$0.12 & m   & \leftthumbsup$^{NEW}$       	\\
		  Sz~51   & 13:01:58.94 & $-$77:51:21.74 & 13.47$\pm$0.02$^\ast$& 12.38$\pm$0.01$^\ast$& 11.80$\pm$0.02 & 12.46$\pm$0.09 & 13.37$\pm$0.13 & 12.55$\pm$0.09 & 12.29$\pm$0.08 & a   & $\bigstar$  	        	\\
		 CM~Cha   & 13:02:13.49 & $-$76:37:57.68 & 12.93$\pm$0.01	& 11.84$\pm$0.01       & 11.18$\pm$0.01 & 12.15$\pm$0.10 & 12.57$\pm$0.13 & 11.55$\pm$0.09 & 11.62$\pm$0.08 & b   & $\bigstar$ \leftthumbsup    	\\
		   C~50   & 13:02:22.82 & $-$77:34:49.51 & 17.78$\pm$0.01	& 15.49$\pm$0.02       & 14.53$\pm$0.01 & 17.78$\pm$0.10 & 18.19$\pm$0.20 & 15.48$\pm$0.09 & 14.93$\pm$0.08 & f,g & $\bullet$ \leftthumbsup     	\\
	IRAS~12589-7646   & 13:02:47.73 & $-$77:02:46.32 &   S  		&	 S	       &    S		& 12.72$\pm$0.09 & 12.92$\pm$0.13 &   S 	   &   S	    & i   & $\bullet$		        	\\
      RXJ1303.1-7706   	  & 13:03:04.46 & $-$77:07:02.75 & 12.27$\pm$0.01$^\ast$& 11.17$\pm$0.01$^\ast$& 10.80$\pm$0.01 & 12.34$\pm$0.09 & 12.08$\pm$0.13 & 11.29$\pm$0.09 & 11.16$\pm$0.08 & c   & $\bigstar$ \leftthumbsup    	\\
		   C~51   & 13:03:09.04 & $-$77:55:59.52 & 16.88$\pm$0.01	& 14.47$\pm$0.01       & 13.46$\pm$0.02 & 16.82$\pm$0.10 & 17.16$\pm$0.13 & 14.39$\pm$0.09 & 14.03$\pm$0.10 & f   & $\bullet$ \leftthumbsup     	\\
	      ChaII~376   & 13:03:12.45 & $-$76:50:50.82 & 16.31$\pm$0.01	& 15.45$\pm$0.01       & 15.35$\pm$0.05 & 16.40$\pm$0.10 & 16.42$\pm$0.13 &      NO        &      NO        & j   & $\bullet$ \leftthumbsdown   	\\
    WFI\,J13031615-7629381& 13:03:16.15 & $-$76:29:38.15 & 16.42$\pm$0.01	& 14.41$\pm$0.03       & 13.68$\pm$0.03 & 16.30$\pm$0.10 & 16.67$\pm$0.15 &	 NO	   &      NO 	    & m   & \leftthumbsup$^{NEW}$       	\\
	  ISO-CHA\,II\,98a& 13:03:25.85 & $-$77:01:48.36 & 17.09$\pm$0.02	& 16.42$\pm$0.04       & 16.24$\pm$0.03 & 17.53$\pm$0.10 & 17.27$\pm$0.11 & 16.83$\pm$0.09 & 17.17$\pm$0.24 & g   & $\bullet$ \ding{68} \leftthumbsdown \\
	  ISO-CHA\,II\,98b& 13:03:26.10 & $-$77:01:48.40 & 17.09$\pm$0.02	& 16.37$\pm$0.01       & 16.18$\pm$0.02 & 17.47$\pm$0.10 & 17.24$\pm$0.11 & 17.68$\pm$0.43 & 16.99$\pm$0.22 & g   & $\bullet$ \ding{68} \leftthumbsdown \\
	  ISO-CHA\,II\,110& 13:04:18.96 & $-$76:53:59.96 & 14.52$\pm$0.01	& 14.03$\pm$0.01       & 14.03$\pm$0.01 & 14.59$\pm$0.09 & 14.58$\pm$0.13 &	 NO	   &	  NO	    & g   & $\bullet$ \leftthumbsdown 		\\
		  Hn~22   & 13:04:22.78 & $-$76:50:05.86 & 13.51$\pm$0.01$^\ast$& 12.68$\pm$0.01$^\ast$& 12.17$\pm$0.02 & 13.23$\pm$0.09 & 13.85$\pm$0.13 &	NO	   &	  NO	    & b   & $\bigstar$ \ding{68}       		\\
	 	  Hn~23   & 13:04:24.06 & $-$76:50:01.39 & 12.23$\pm$0.01$^\ast$& 11.33$\pm$0.08$^\ast$&     S          & 12.11$\pm$0.09 & 12.34$\pm$0.13 &	NO	   &	  NO	    & b   & $\bigstar$ \ding{68} \leftthumbsup  \\
		  Sz~52   & 13:04:24.84 & $-$77:52:30.11 & 16.89$\pm$0.01	& 14.94$\pm$0.01       & 14.12$\pm$0.01 & 17.00$\pm$0.10 & 17.35$\pm$0.13 & 14.84$\pm$0.09 & 14.43$\pm$0.08 & a   & $\bigstar$ \leftthumbsup  		\\
		  Hn~24   & 13:04:55.74 & $-$77:39:49.21 & 13.30$\pm$0.06$^\ast$& 11.95$\pm$0.02$^\ast$& 11.50$\pm$0.02 & 13.49$\pm$0.09 & 13.43$\pm$0.14 & 12.19$\pm$0.09 & 12.01$\pm$0.08 & b   & $\bigstar$ \leftthumbsup    	\\
		  Hn~25   & 13:05:08.51 & $-$77:33:42.66 & 16.11$\pm$0.01	& 14.17$\pm$0.02       & 13.36$\pm$0.01 & 16.45$\pm$0.12 & 16.61$\pm$0.13 & 14.40$\pm$0.09 & 14.01$\pm$0.09 & b   & $\bigstar$ \leftthumbsup    	\\
		  Sz~53   & 13:05:12.66 & $-$77:30:52.56 & 15.58$\pm$0.02	& 13.86$\pm$0.02       & 13.15$\pm$0.01 & 15.58$\pm$0.09 & 15.90$\pm$0.18 & 15.16$\pm$0.12 & 14.47$\pm$0.08 & a   & $\bigstar$ \leftthumbsup    	\\
		  Sz~54   & 13:05:20.80 & $-$77:39:01.48 & 11.58$\pm$0.02$^\ast$& 10.61$\pm$0.06$^\ast$& 10.47$\pm$0.43 & 11.28$\pm$0.10 & 11.57$\pm$0.13 & 10.80$\pm$0.09 & 11.05$\pm$0.15 & a   & $\bigstar$ \leftthumbsup    	\\
		  Sz~55   & 13:06:30.49 & $-$77:34:00.12 & 16.37$\pm$0.01	& 14.70$\pm$0.06       & 14.06$\pm$0.47 & 15.65$\pm$0.09 & 16.96$\pm$0.13 & 14.61$\pm$0.09 & 14.53$\pm$0.08 & a   & $\bigstar$ \leftthumbsup    	\\
		  Sz~56   & 13:06:38.70 & $-$77:30:35.39 & 15.41$\pm$0.01$^\ast$& 13.47$\pm$0.02$^\ast$& 12.71$\pm$0.44 & 15.41$\pm$0.09 & 15.75$\pm$0.13 & 13.67$\pm$0.09 & 13.15$\pm$0.08 & a   & $\bigstar$ \leftthumbsup    	\\
		  Sz~57   & 13:06:56.56 & $-$77:23:09.46 & 15.97$\pm$0.01	& 13.65$\pm$0.01       & 12.59$\pm$0.41 & 15.71$\pm$0.09 & 16.23$\pm$0.13 & 13.74$\pm$0.09 & 13.07$\pm$0.08 & a   & $\bigstar$ \leftthumbsup    	\\
		  Sz~58   & 13:06:57.35 & $-$77:23:41.46 & 14.63$\pm$0.01	& 13.16$\pm$0.01       & 12.53$\pm$0.43 & 14.44$\pm$0.09 & 14.58$\pm$0.13 & 12.98$\pm$0.09 & 12.78$\pm$0.08 & a   & $\bigstar$ \leftthumbsup    	\\
		  Sz~59   & 13:07:09.23 & $-$77:30:30.24 & 13.42$\pm$0.02$^\ast$& 12.08$\pm$0.01$^\ast$& 11.51$\pm$0.43 & 12.96$\pm$0.10 & 13.39$\pm$0.13 & 12.57$\pm$0.09 & 12.50$\pm$0.08 & a   & $\bigstar$ \leftthumbsup    	\\
		   C~62   & 13:07:18.04 & $-$77:40:53.00 & 18.60$\pm$0.01	& 16.10$\pm$0.01       & 15.00$\pm$0.44 & 18.42$\pm$0.11 & 18.97$\pm$0.17 & 16.23$\pm$0.09 & 15.60$\pm$0.08 & f,l & $\bullet$ \leftthumbsup     	\\
    WFI\,J13071960-7655476& 13:07:19.60 & $-$76:55:47.64 & 18.33$\pm$0.01	& 16.49$\pm$0.02       & 15.77$\pm$0.02 & 18.12$\pm$0.10 & 18.57$\pm$0.14 &	  NO	   &        NO	    & m   & \leftthumbsup$^{NEW}$       	\\
		 Sz~60W   & 13:07:22.30 & $-$77:37:22.62 & 14.88$\pm$0.02$^\ast$& 13.45$\pm$0.02$^\ast$& 13.02$\pm$0.46 & 14.60$\pm$0.09 & 15.01$\pm$0.13 & 13.42$\pm$0.09 & 13.23$\pm$0.07 & a   & $\bigstar$ \ding{68} \leftthumbsup  \\
		 Sz~60E   & 13:07:23.33 & $-$77:37:23.20 & 15.32$\pm$0.02$^\ast$& 13.60$\pm$0.01$^\ast$& 12.94$\pm$0.46 & 15.01$\pm$0.09 & 15.69$\pm$0.13 & 13.98$\pm$0.09 & 13.61$\pm$0.08 & a   & $\bigstar$ \ding{68}          	\\
		  Hn~26   & 13:07:48.50 & $-$77:41:21.73 & 16.07$\pm$0.01	& 14.31$\pm$0.01       & 13.54$\pm$0.46 & 15.89$\pm$0.11 & 16.25$\pm$0.13 & 14.41$\pm$0.09 & 14.13$\pm$0.08 & b   & $\bigstar$ \leftthumbsup      	\\
		  Sz~61   & 13:08:06.33 & $-$77:55:05.05 & 13.69$\pm$0.02$^\ast$& 12.38$\pm$0.01$^\ast$& 11.48$\pm$0.42 & 12.69$\pm$0.09 & 13.47$\pm$0.13 & 12.27$\pm$0.09 & 12.14$\pm$0.08 & a   & $\bigstar$ \leftthumbsup      	\\
		   C~66   & 13:08:27.19 & $-$77:43:23.41 & 18.91$\pm$0.01	& 16.51$\pm$0.01       & 15.44$\pm$0.44 & 19.24$\pm$0.14 & 19.72$\pm$0.22 & 16.76$\pm$0.09 & 16.03$\pm$0.08 & f,l & $\bullet$ \leftthumbsup       	\\
      IRAS~F13052-7653NW  & 13:09:09.81 & $-$77:09:43.52 & 15.37$\pm$0.01	& 14.04$\pm$0.02       & 13.52$\pm$0.44 & 15.59$\pm$0.09 & 16.11$\pm$0.13 & 14.02$\pm$0.09 & 13.84$\pm$0.08 & e   & $\bullet$ \ding{68}	          	\\
      IRAS~F13052-7653S   & 13:09:10.67 & $-$77:09:46.84 & 13.49$\pm$0.01	& 12.57$\pm$0.01       & 12.38$\pm$0.47 & 13.09$\pm$0.09 & 13.04$\pm$0.13 & 12.69$\pm$0.09 & 12.77$\pm$0.08 & e   & $\bullet$ \ding{68}	          	\\
      IRAS~F13052-7653N   & 13:09:10.98 & $-$77:09:44.14 & 13.47$\pm$0.01	& 12.42$\pm$0.02       & 12.07$\pm$0.45 & 13.51$\pm$0.09 & 13.74$\pm$0.13 & 12.54$\pm$0.09 & 12.47$\pm$0.08 & e   & $\bullet$ \ding{68}	          	\\
		  Sz~62   & 13:09:50.44 & $-$77:57:23.94 & 14.03$\pm$0.02$^\ast$& 12.56$\pm$0.02$^\ast$& 11.66$\pm$0.43 &      NO	 &	  NO	  &     NO	   &	  NO	    & a   & $\bigstar$ \leftthumbsup      	\\
		  Sz~63   & 13:10:04.12 & $-$77:10:44.62 & 14.66$\pm$0.01	& 13.21$\pm$0.01       & 12.66$\pm$0.44 & 14.40$\pm$0.09 & 15.09$\pm$0.13 & 13.25$\pm$0.09 & 13.03$\pm$0.08 & a   & $\bigstar$ \leftthumbsup      	\\
 2MASS13102531-7729082    & 13:10:25.26 & $-$77:29:08.70 & 16.49$\pm$0.01  	& 15.36$\pm$0.05       & 14.95$\pm$0.48 & 16.49$\pm$0.10 & 16.49$\pm$0.13 & 15.54$\pm$0.09 & 15.43$\pm$0.08 & i   & $\bullet$ \leftthumbsdown     	\\
 2MASS13110329-7653330    & 13:11:03.27 & $-$76:53:32.89 & 14.50$\pm$0.01  	& 13.85$\pm$0.00       & 13.68$\pm$0.01 & 14.62$\pm$0.09 & 14.57$\pm$0.13 &	 NO	   &      NO 	    & i   & $\bullet$ \leftthumbsdown     	\\
 2MASS13125238-7739182    & 13:12:52.37 & $-$77:39:18.58 & 15.36$\pm$0.01  	& 13.62$\pm$0.02       & 12.97$\pm$0.47 & 15.57$\pm$0.09 & 15.78$\pm$0.13 & 13.83$\pm$0.09 & 13.51$\pm$0.08 & i   & $\bullet$ \leftthumbsup   		\\
		  Sz~64   & 13:14:03.83 & $-$77:53:07.48 & 15.82$\pm$0.01	& 14.01$\pm$0.01       & 13.16$\pm$0.45 & 15.25$\pm$0.09 & 16.29$\pm$0.13 & 14.40$\pm$0.09 & 13.65$\pm$0.08 & a   & $\bigstar$ \leftthumbsup  		\\
\end{longtable}
\footnotesize{Main references: a) \citet{Hug92}; b) \citet{Har93}; c) \citet{Alc95}; d) \citet{Cov97}; e)\citet{Alc00}; f) \citet{Vuo01}; 
g) \citet{Per03}; h) \citet{Bar04}; i) \citet{You05}; j) \citet{Lop05}; k) \citet{Alc06}; l) \citet{All06}; m) This work.}  \\

\footnotesize{Notes: 
\begin{itemize}
\item $\bigstar$ = Confirmed member 
\item $\bullet$ = Previously known candidate
\item \leftthumbsup  = Object recovered by our selection criteria
\item \leftthumbsup$^{NEW}$  = Member candidate firstly identified in this work
\item \leftthumbsdown  = Object rejected on the basis of our selection criteria
\item \ding{80} = Veiled object
\item \ding{74} = Embedded object
\item \ding{68} = Component of a visual binary or multiple system
\item S = Saturated object
\item ND = Object not detected 
\item NO = Object not observed 
\end{itemize}
}		   
\end{landscape}

\newpage

\topmargin -1.7cm
\twocolumn
\pagestyle{headings}

\newpage
\clearpage

%---------------------------------- CMDs-------------------
%-------------------------------------------------------------

\newpage
\clearpage

\appendix

\section{Flux at $z=0$ \label{fluxZ} }
   
In order to derive the flux at zero magnitude in the $z$ band we 
used the following procedure:

\begin{enumerate}

\item 
The spectrophotometry of SA~98~653 is available from \citet{Gut88}
in the form of monochromatic instrumental magnitudes, covering 
a wavelength range from 3200 to 8000 \AA, in steps of 40 \AA~ 
for $\lambda \le 5400$ \AA~ and 80 \AA~ for $\lambda > 5400 $\AA. 
Since the $z$ band is centered at 9648 \AA~ (FWHM=616~\AA),
we need the spectrophotometry in a wider wavelength range.
We have thus extrapolated the SED of SA~98~653, up to 16000~\AA, 
by using the Kurucz grid of stellar 
atmosphere models \citep{Kur79} in the following way. We first 
dereddened the SED of SA~98~653 using 
the canonical absorption-to-reddening ratio R$_{\rm V}$=3.2, as 
suggested by \citet{Che01} for the SA~98 field. Assuming that 
SA~98~653 is a main sequence star, as indicated by its 
spectrophotometry and broad-band photometric indexes, we derived 
its distance by comparing its observed $V$ magnitude with the 
expected absolute visual magnitude ($M_{V}=0.7$, Gray 1992) 
for an A0-type main sequence star. 
Assuming an error of two spectral sub-classes on SA~98~653 
spectral type, we constrained its distance between $300$ and $450$ 
pc, i.e. 0.30~mag$<$ A$_{\rm V}$ $<$0.45~mag. Note that the 
extinction of the nearby star SA~98-667 (spectral type B9V) 
is A$_{\rm V}$=0.33 calculated from its Hipparcos parallax 
\citep{Che01}, which is consistent with the range of values 
allowed for SA~98~653. 
The interstellar extinction coefficients at each wavelength were 
derived by using the extinction curve by \citet{Car89}. 
We then performed a two-parameters least-square fit to the observed 
spectrophotometry of SA~98~653 by taking as free parameters the 
extinction (in the range 0.30~mag$<$ A$_{\rm V}$ $<$0.45~mag in 
steps of 0.01~mag) and the effective temperature of the model 
spectrum. 
Both the Kurucz models and the dereddened SED of SA~98~653 
were normalised to the flux at 5556\,\AA. 
A minimum $\chi^2$ for a model with T$_{\rm eff}=11000$~K, 
$\log g$=4 and A$_{\rm V}$=0.33 was derived (Fig.~\ref{SA98653}). 
We then used this synthetic spectrum reddened with A$_{\rm V}$=0.33 to extrapolate 
the spectrophotometry of SA~98~653 in the near-IR.

%----------------------------------SA 98 653 -------------------
\begin{figure} 
\resizebox{\hsize}{!}{\includegraphics[width=10cm,height=7cm]{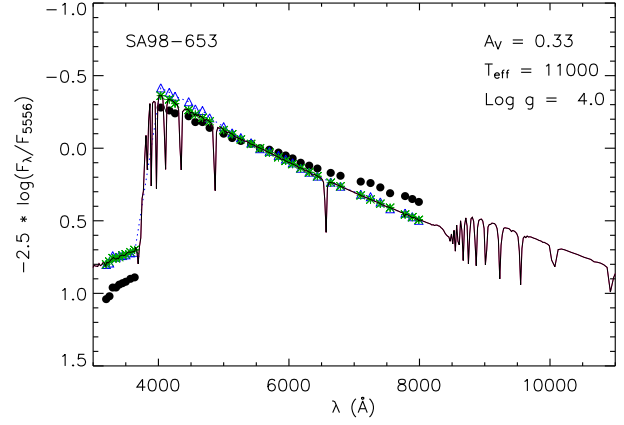}}
\caption{Observed (dots) and dereddened (open triangles) spectral energy distribution 
of SA~98~653 normalised to the flux at 5556\,\AA. The best fitting model by \citet{Kur79} is over-plotted (continuous line).}
\label{SA98653}
\end{figure}
%-------------------------------------------------------------------

\item 
The absolute Earth flux at 5556 \AA~ of a star with magnitude $V^*$ 
calibrated with Vega, the primary photometric standard, is given by \citep{Gra92}:

\begin{equation}
log(F^*_{5556})=-0.400 \cdot V^*-8.451
\label{eqF}
\end{equation}

in units of erg $\cdot$ cm$^{-2}$ $\cdot$ s$^{-1}$ $\cdot$ ~\AA$^{-1}$.

When this expression is used for stars whose effective temperature 
is different from Vega (T$_{\rm eff}=9790$~K), as in the case of SA~98~653, 
a small systematic error is introduced because the effective wavelength 
of the $V$ band is 5480 \AA, i.e. 76 \AA~ shorter than the reference 
wavelength (5556 \AA). An adequate correction for this is given by 
the following colour term \citep{Gra92}:

\begin{equation}
log \left (\frac{F^*_{5556}}{F^*_{5480}} \right )= -0.006+0.018 \cdot (B-V)^* 
\label{eqV}
\end{equation}

Considering the dereddened visual magnitude of SA~98~653 and its 
intrinsic $(B-V)$, and applying Eq.~\ref{eqV} and Eq.~\ref{eqF} 
respectively, we obtain:

$$ 
V^{corr} _{SA~98~653} = 9.24
$$

and

$$
log[F^{5556} _{SA~98~653}]= -12.1465
$$

We now can calculate the flux of SA~98~653 at any wavelength with 
respect to the flux at 5556 \AA~ as: 

\begin{equation} 
F^{synt} _{SA~98~653} (\lambda) =F^{5556} _{SA~98~653} \cdot 10^{-0.4 \cdot mag^{synt}_{SA~98~653}(\lambda)}
\end{equation}

where $mag^{synt} _{SA~98~653}$ is the synthetic magnitude of SA~98~653 
at each wavelength normalised to the magnitude at 5556 \AA, determined 
in the previous step.

\item 
We finally determined the synthetic flux of SA~98~653 in the $z$ band 
by integrating its synthetic spectrum $F^{synt} _{SA~98~653} (\lambda)$ 
in the $z$ filter transmission curve T$_{\rm \lambda}$, i.e. the product of 
the detector quantum efficiency and the filter transmission curve:

\begin{equation} 
F^z _{SA~98~653}=\frac{1}{A_z} \cdot \int_{\lambda_1} ^{\lambda_2}{F^{synt} _{SA~98~653}(\lambda)\cdot T_{\lambda} \cdot d\lambda}
\end{equation}

where:

\begin{equation} 
A_z=\int_{\lambda_1} ^{\lambda_2} {T_{\lambda} \cdot d\lambda}
\end{equation}

We obtained: 

$$
F^z _{SA~98~653}=1.62 \cdot 10^{-13}  \hspace{0.5cm} erg \cdot cm^{-2} \cdot s^{-1} \cdot \AA^{-1}
$$

$F^{z} _{SA~98~653}$ is the flux in the $z$ filter of an A0 type star 
with visual magnitude $V=9.539$ and $z=I_C=9.522$. What we need, in order 
to obtain an absolute flux calibration for the $z$ band, is the flux of 
an A0-type star with $V=0$ and $z=0$ ($F^z _{A0V}$). 
Since SA~98~653 is an A0-type star and we have defined its intrinsic 
colour as $(I_C-z)=0$, we have:

\begin{equation} 
z_{SA~98~653} \stackrel{\text{def}}{=} I_{SA~98~653} = 9.522
\end{equation}

Thus, we derive $F^z _{A0V}$ as:

\begin{equation}
F^z _{A0V} = F^z _{SA~98~653} \cdot 10^{(0.4 \cdot z_{SA~98~653})}
\end{equation}

By considering the uncertainties on magnitude and interstellar 
extinction measurements, we obtained:

$$
F^z _{A0V} = (8.4 \pm 0.1) \cdot 10^{-10} \hspace{0.25cm} {\rm erg} \hspace{0.1cm} {\rm cm}^{-2} {\rm s}^{-1} {\rm \AA}^{-1}
$$ 
$$ 
\hspace{-2.3cm}       = (2608 \pm 31)\ {\rm Jy}
$$

\end{enumerate}

A slightly smaller absolute flux calibration constant for the 
WFI $z$ band has been estimated by F. Com\'eron (private communication) 
using the absolute flux calibration of Vega obtained by \citet{Boh04} 
with the Space Telescope Imaging Spectrograph (STIS). Using these 
data we derive a value of 7.2 $\cdot$ 10$^{-10}$ erg $\cdot$ cm$^{-2}$ $\cdot$ s$^{-1}$ $\cdot$ \AA$^{-1}$. 
However, for the sake of homogeneity, we prefer to use the value we derive above.

\section{Isochrones \label{isocs}}

In order to transform the colours and magnitudes of the isochrones
reported by \citet{Bar98} and \citet{Cha00} into the WFI-Cousins system 
we determined first the WFI synthetic magnitudes, $m_{\Delta \lambda}$:

\begin{equation}
m_{\Delta \lambda} = -2.5 \cdot log_{10} (f_{\Delta \lambda}) + C_{\Delta \lambda} 
\label{EqMag}
\end{equation}

where $C_{\Delta \lambda}$ is the absolute calibration constant of the 
photometric system, tied to the Earth flux of an A0-type star with
magnitude $V=0$ (see Tab.~\ref{tab:flux0}), and $f_{\Delta \lambda}$ 
is the observed flux.

In order to compute the  expected flux of a PMS star or a young BD in 
the pass-band $\Delta \lambda$ of a given filter we proceeded as follows. 
We first determined the transmission curve, $T_{\lambda}$,
i.e. the product of the detector quantum efficiency and the given 
filter transmission curve\footnote{Available at: http://www.ls.eso.org/lasilla/sciops/2p2/E2p2M/WFI/ filters.}. 
We then computed:

\begin{equation}  
f_{\Delta \lambda} = \frac{1}{A_{\Delta \lambda}} 
\cdot \int_{\lambda_1} ^{\lambda_2} F_{\lambda} \cdot T_{\lambda} \cdot d\lambda
\end{equation}

where

\begin{equation}   
A_{\Delta \lambda} = \int_{\lambda_1} ^{\lambda_2} T_{\lambda} \cdot d\lambda
\end{equation}

and $F_{\lambda}$ is the absolute flux of a given PMS star or BD. 
For this purpose we used the synthetic low-resolution spectra for 
low-mass stars by \citet{Hau99} calculated with their NextGen 
model-atmosphere code. For simulating very cool objects (i.e. T$_{\rm eff}<$3000~K) 
we used the StarDusty and BD-Dusty atmosphere models by \citet{All00}, which include dust opacity.

%---------------------------------- Synthetic Spectra + WFI filter-------------------
\begin{figure} 
\resizebox{\hsize}{!}{\includegraphics[width=12cm,height=13cm]{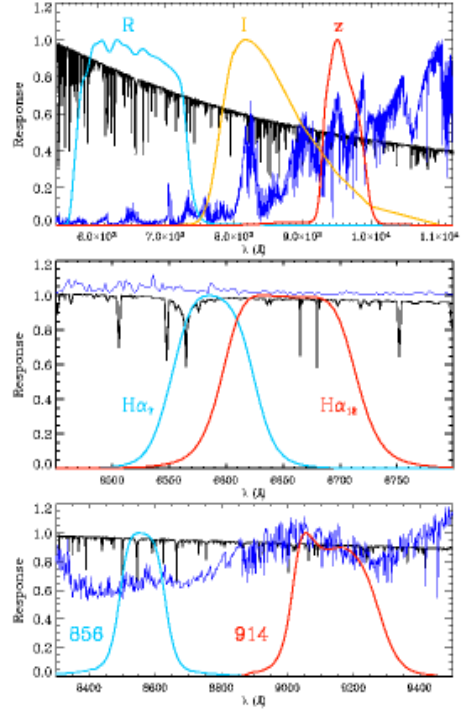}}
\caption{The $R$, $I$, $z$ (upper panel), $H\alpha_7$, $H\alpha_{12}$ (central panel), 
856 and 914 nm (lower panel) WFI+ESO2.2m transmission bands. Examples of two 
normalised NextGen spectra \citep{Hau99} with T$_{\rm eff}$=2500~K and 
T$_{\rm eff}$=6000~K are over-plotted.}
\label{fig:filter}
\end{figure}
%-----------------------------------------------------------------------------

%---------------------------------------Landolt Cousins-WFI------------------- 
\begin{figure}   
\centering
\includegraphics[width=7cm,height=9cm]{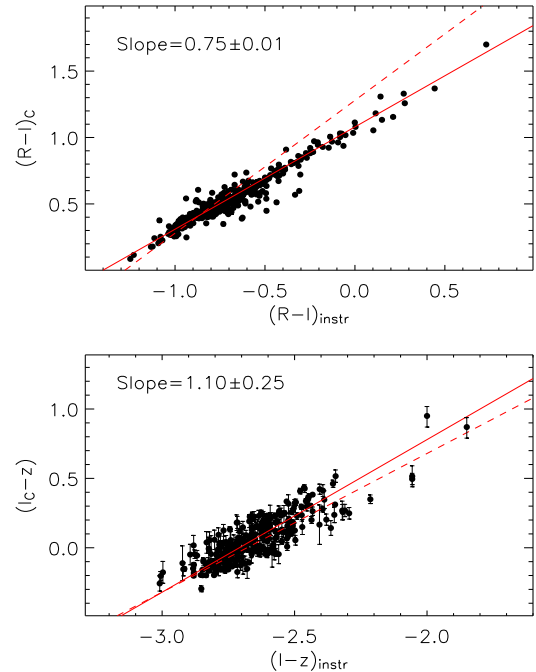}
\caption{The Cousins colours as a function of the WFI instrumental colours 
for the standard stars in the Landolt's field SA~98. The continuous lines 
display the best fit of the data while the dashed lines have slope equal 
to 1 (i.e. no correction).}
\label{fitRIz}
\end{figure}
%-----------------------------------------------------------------------------

As an example of our procedure we show in 
Fig.~\ref{fig:filter} (upper panel) two model spectra and the $R$, 
$I$ and $z$ WFI+ESO2.2m transmission curves over-plotted. However, 
the flux $f_{\Delta \lambda}$ that we obtained in this way is the 
one at the surface of the star. The observed flux at any distance 
(d) is given by:

\begin{equation}
f_{\Delta \lambda} (d) = f_{\Delta \lambda} \cdot \left (\frac{R^\star}{d}  \right )^2
\end{equation}

where R$^{\star}$ is the object radius. The values for R$^{\star}$ 
can be computed from theoretical PMS evolutionary tracks, which in 
our case are those by \citet{Bar98} for low-mass stars and those 
by \citet{Cha00} for sub-stellar objects (i.e. M$\lesssim 0.1 M_{\odot}$). 
Finally, setting d=10~pc and using Eq.~\ref{EqMag}, one can derive 
the absolute magnitudes for a given T$_{\rm eff}$, age and R$^{\star}$ 
and hence, determine the isochrones in the CMDs.

Note that, consistently with the $RIz$ photometric calibration of 
the sources in Cha~II, the absolute calibration constants used 
in Eq.~\ref{EqMag} are those of the WFI-Cousins photometric system 
(see Tab.~\ref{tab:flux0}). However, a non-negligible difference 
between the $I$-WFI and the $I$-Cousins pass-bands arises from the 
fact that the red cut-off of the $I$-WFI filter is defined by the 
CCD response and it is centered at a redder wavelength (0.85$\mu$m) 
with respect to the standard Cousins $I$ filter (0.79$\mu$m). 
In order to take into account this difference in the determination 
of the theoretical isochrones, we have exploited the relations 
between the WFI-Cousins ($R-I$)$_C$ and ($I_C-z$) standard colours 
and the WFI instrumental colours for the standard stars in the 
Landolt's fields. 
By a least-square linear fitting (Fig.~\ref{fitRIz}), we derive the 
coefficients which allow us to correct the colours of the computed 
isochrones:

\begin{equation}
(R-I)_{synt} = 0.75 \cdot (R-I)_{instr}
\end{equation}

\begin{equation}
(I-z)_{synt} = 1.10 \cdot (I-z)_{instr}
\end{equation}

In this way we were able to report in the WFI-Cousins photometric 
system the isochrones originally calculated by \citet{Bar98} and 
\citet{Cha00} in the Cousins system of Bessel \citep{Bes90}. 
This allows us to use the isochrones in a photometrically 
homogeneous way. 

\begin{figure}   
\centering
\resizebox{\hsize}{!}{\includegraphics[width=9cm,height=7cm]{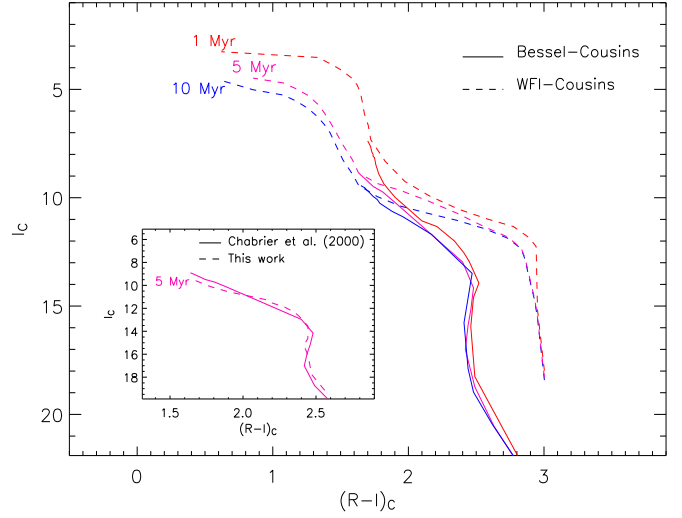}}
\caption{Theoretical isochrones by \citet{Bar98} and \citet{Cha00} for the 
Cousins photometric system of Bessel (continuous line) and for the 
WFI-Cousins system (dashed lines, this work). 
In the smaller panel the 5 Myr isochrone by \citet{Cha00} (continuous line) 
is compared with that reproduced by our algorithm when using the same 
Bessel filters (dashed line).}
\label{isoc_Bes}
\end{figure}

In Fig.~\ref{isoc_Bes} the 1, 5 and 10 Myr isochrones by \citet{Cha00} 
are compared with those derived by us in the WFI-Cousins photometric 
system. Note the large difference, in particular for the coolest 
objects, i.e. $(R-I)_C >$1.7, where the use of the isochrones in 
the Bessel-Cousins system would have produced many spurious member 
candidates.

In order to check our procedure, we used the standard filters of 
\citet{Bes90} and we were successful in reproducing the isochrones 
presented by \citet{Cha00} (see the smaller panel in Fig.~\ref{isoc_Bes}).

%---------------------------------------Bessel-WFI------------------- 
%-----------------------------------------------------------------------------

\section{Comments on individual objects}
\label{comm}

\subsection*{IRAS~12416-7703, IRAS~12448-7650, IRAS~F12488-7658, IRAS~12556-7731, IRAS~12589-7646}
\label{cand_sat}

These objects were selected as member candidates of Cha~II 
by \citet{You05} on the basis of MIPS@Spitzer observations. 
IRAS~F12488-7658/C~13 was also previously proposed as a young object by \citep{Vuo01}. 
We cannot conclude whether our selection criteria are satisfied
for these objects because they are saturated in at least one of the 
three broad-bands $RIz$. However, their ($H\alpha_{12}-H\alpha_{7}$) 
indices suggest H$\alpha$ emission (see Tab.~\ref{tab:phot})  at $\sim$2$\sigma$ level and their 
location on the ($J-H$) vs. ($H-K$) diagram is consistent with 
the PMS stars locus (see Sec.~\ref{sel_cand} and Fig.~\ref{fig:JHK_diag}).

\subsection*{ChaII~304, ChaII~305, ChaII~376}

ChaII~376 was the only H$\alpha$ emitter detected by \citet{Lop05}
in their WFI pointing.  %in Cha~II. 
They also proposed that their candidates ChaII~304 and ChaII~305 are two 
low-mass BDs or planetary-mass objects close to the Deuterium burning limit. 
These two objects, however, do not satisfy our selection criteria. 
The early classification of these objects as possible Cha~II members 
by \citet{Lop05}, though based on WFI images taken with the same set of 
filters as in our survey, is probably not accurate   %incorrect 
because the isochrones in the Cousins system of Bessel \citep{Cha00} used 
for the selection are inadequate for the WFI-Cousins photometric system
(see Appendix~\ref{isocs}). From their dereddened ($m_{856}-m_{914}$) colour 
index we estimate an effective temperature of 2800~K and 3000~K for ChaII~304 
and ChaII~305, respectively. Assuming these temperature values, we find 
that the location of the two objects on the HR diagram would be 
inconsistent with membership to Cha\,II. 
We found that the optical colours of ChaII~376 are inconsistent 
with membership, as also asserted by \citet{Lop05}, and we do not
detect any significant H$\alpha$ emission for this object. 
Moreover, its $R$-band magnitude (16.31~mag) is remarkably brighter 
than the one reported by \citet{Lop05} (18.38~mag). Since the 
magnitude of this object in the other bands are in fair agreement 
with ours, we think that the $R$-band magnitude reported in 
\citet{Lop05} may be incorrect. This may explain why ChaII~376 
was selected as their only H$\alpha$ emitter.

\subsection*{Sz~46}

This is one of the 19 PMS stars originally identified by \citet{Sch77} 
in Cha~II. Subsequent photometric observations by \citet{Hug92} showed 
that the object is a visual triple system. The H$\alpha$ emission 
star originally discovered by \citet{Sch77}, Sz~46N, has two companions, 
namely Sz~46NW and Sz~46S; both are bluer than Sz~46N and are probably 
unrelated to it \citep{Hug92}.
Based on our analysis, both Sz~46NW and Sz~46S have optical colours 
inconsistent with membership to the Cha~II cloud, their temperature is 
well above 4000~K according to their ($m_{856}-m_{914}$) index and
the ($H\alpha_{12}-H\alpha_{7}$) colour suggests no H$\alpha$ emission. 
Thus, Sz~46NW and Sz~46S are most probably not PMS stars and hence, not 
physically associated with Sz~46N.

\subsection*{IRAS~12533-7632, IRAS~F13052-7653}

These objects were detected as ROSAT X-rays sources \citep{Alc00} 
and proposed as possible members of Cha~II by 
different authors \citep{Pru92,Lar98,You05}. 

IRAS~12533-7632 appears on our images as a point-like object 
associated with a nebulosity, consistently with what reported 
in \citet{Pru92}. However, it was not detected in our $z$ and 
H$\alpha$ images; its $RI$ magnitudes and 2MASS near-IR colours 
are inconsistent with those expected for a PMS star. 

In \citet{You05} IRAS~F13052-7653 is reported as CHIIXR~60 which 
is a non-existent identifier. 
This must be a typo because IRAS~F13052-7653 is identified with 
the ROSAT source CHIIXR~37 \citep{Alc00}.  
The optical counterpart of IRAS~F13052-7653 appears as a visual 
triple system \citep{Lar98} whose components are designated as 
IRAS~F13052-7653N, IRAS~F13052-7653NW and IRAS~F13052-7653S in 
Tab.~\ref{tab:phot}. 
None of the three components was recovered by our selection criteria, 
most likely because their colours, both in optical and and near-IR, 
may be contaminated. However, two of them, namely 
IRAS~F13052-7653N and IRAS~F13052-7653NW, fall just below the 
10 Myr isochrone and their ($H\alpha_{12}-H\alpha_{7}$) indices 
suggest H$\alpha$ emission above 3$\sigma$ level (see Tab.~\ref{tab:phot}); moreover,
their near-IR colours are also consistent with those expected 
for PMS stars. IRAS~F13052-7653S is not detected in 2MASS; 
its optical colours are inconsistent with membership to Cha~II and 
its H$\alpha$ colour does not indicate significant emission.

\subsection*{IRAS~F12571-7657}

This object was detected in X-rays with ROSAT \citep{Alc00}. 
Its optical and near IR colours are consistent with those of 
a PMS star with strong IR excess (see also Young et al. 2005). 
Previous authors \citep{Vuo01, All06} report an extinction value 
significantly higher than what we determine from the \citet{Cam99} 
extinction map. 
Given the strong IR excess, our procedure for the temperature 
estimate provides unreliable results. Therefore, it may be a 
deeply embedded source or a highly reddened PMS star.

\subsection*{C~62, C~66}     

These sources have been firstly detected by DENIS and 
proposed as young stellar object candidates of Cha~II 
by \citet{Vuo01} based on their IR colours and, 
more recently, by \citet{You05} and \citet{All06} 
on the basis of IRAC@Spitzer and MIPS@Spitzer observations
respectively. From the effective temperature and A$_{\rm V}$ 
estimates provided in Tab.~\ref{tab:par}, and calculating 
their luminosities as explained in Sec.~\ref{sec:par},  
we estimate a mass of 0.08 and 0.06~M$_{\odot}$ 
for C~62 and C~66, respectively. The age we derive 
for both objects using the \citet{Bar98} tracks is 
around 2 Myr. These estimates of mass and age are 
fairly consistent with those derived by \citet{All06}. 
From their H$\alpha$ index we also notice strong line 
emission ($\sim$3$\sigma$ level) in both objects (see Tab.~\ref{tab:par}).

\subsection*{ISO-CHA\,II\,29}

This source has been detected in the near and mid-IR 
by \citet{Per03}, but it was first revealed as an X-ray 
emitting source, CHIIXR3, by \citet{Alc00}. 
The association of this object with IRAS~12551-7657 is 
still unclear. Based on these evidences and 
the effective temperature estimated by us (Tab.~\ref{tab:par}), 
ISO-CHA\,II\,29 is proposed to be a PMS star in Cha~II 
with a mass of about 1~M$_{\odot}$.

\subsection*{ISO-CHA\,II\,73,  ISO-CHA\,II\,98, ISO-CHA\,II\,110}

None of these sources, originally identified by \citet{Per03}, 
were selected with MIPS@Spitzer by \citet{You05}.  These sources 
do not satisfy our criteria and their ($H\alpha_{12}-H\alpha_{7}$) 
index suggests no emission in H$\alpha$. 

In our optical images ISO-CHA\,II\,98 appears as a visual binary,
but the colours and magnitudes of its components are inconsistent 
with our selection criteria. 
We stress that ISO-CHA\,II\,98 has been wrongly identified in 
SIMBAD with the ROSAT X-ray source CHIIXR~24 by \citet{Alc00}. 
The optical spectrum of CHIIXR~24 shows the Li\,I 6708\AA~ 
absorption line, but the H$\alpha$ line is well in 
absorption \citep{Alc00}. Being classified as a G7-type star,
CHIIXR~24 is most likely a field object unrelated to Cha~II.

\subsection*{2MASS\,13102531-7729085, 2MASS\,13110329-7653330, 2MASS\,13125238-7739182}

These sources are reported in the list of candidates by
\citet{You05}. While 2MASS\,13102531-7729085 and 2MASS\,13110329-7653330
are rejected by our selection criteria, 2MASS\,13125238-7739182
is selected as a PMS star candidate. 
The former two sources have H$\alpha$ indices indicating no 
H$\alpha$ emission, while the  H$\alpha$  index of the latter 
suggests H$\alpha$ in emission at 1$\sigma$ level. Thus, it is likely that 
2MASS\,13102531-7729085 and 2MASS\,13110329-7653330 are 
field stars unrelated to Cha~II.

\end{document}